\documentclass[twocolumn]{aa} 

\usepackage{graphicx}
\usepackage{txfonts}
\usepackage[colorlinks=True,allcolors=blue]{hyperref}

\usepackage{mathrsfs}
\usepackage{amsbsy}

\usepackage{pifont}
\usepackage{subcaption}
\begin{document}

   \title{Possibility of Concentration of Non-volatile Species near the Surface of Comet 67P/Churyumov-Gerasimenko}

   \subtitle{}

   \author{T. Suzuki\inst{1,2}
          \and
          Y. Shinnaka\inst{3,2}
          \and
          L. Majumdar\inst{4}
          \and
          T. Shibata\inst{2,5,6}
          \and
          Y. Shibaike\inst{7}
          \and
          H. Nomura\inst{7}
          \and
          H. Minamoto\inst{7}
          }
   \institute{Astrobiology Center, Osawa 2-21-1, Mitaka, Tokyo 181-8588, Japan
       \email{taiki.suzuki@nao.ac.jp}
         \and
             National Astronomical Observatory of Japan, 2-21-1 Osawa, Mitaka, Tokyo, 181-8588, JAPAN
         \and
             Laboratory of Infrared High-resolution Spectroscopy (LiH), Koyama Astronomical Observatory, Kyoto Sangyo University, Motoyama, Kamigamo, Kita-ku, Kyoto 603-8555, Japan
          \and      
          School of Earth and Planetary Sciences, National Institute of Science Education and Research, HBNI, Jatni 752050, Odisha, India
              \email{liton@niser.ac.in}  
         \and
             The University of Tokyo, 7-3-1 Hongo, Bunkyo-ku, Tokyo, 113-8654, JAPAN
         \and
             Core Concept Technologies Inc., 11F DaiyaGate Ikebukuro, 1-16-15 Minamiikebukuro, Toshima-ku, Tokyo-to 171-0022, Japan
        \and               
            Department of Earth and Planetary Sciences, Tokyo Institute of Technology, 2-12-1 Ookayama, Meguro-ku, Tokyo, 152-8551, Japan
        }
   \date{Received **, ****; accepted **, ****}

 
  \abstract
{The cometary materials are thought to be the reservoir of primitive materials in the Solar System.
The recent detection of glycine and CH$_3$NH$_2$ by the ROSINA mass spectrometer, in the coma of 67P/Churyumov-Gerasimenko, suggests that amino acids and their precursors may have formed in an early evolutionary phase of the Solar System.}
{We investigate the evolution of cometary interior considering the evaporation process of water, followed by the concentration of non-volatile species.}
{We develop a Simplified Cometary Concentration Model (SCCM) to simulate the evaporation and concentration processes on the cometary surface.
We use 67P/Churyumov-Gerasimenko as the benchmark of SCCM.
We investigate the depth of the layer where non-volatile species concentrate after the numerous passages of perihelion after the formation of the Solar System.}
{We find that the SCCM explains the observed production rates of water and CH$_3$NH$_2$ at 100 comet years.
SCCM results suggest that the non-volatile species would concentrate at depths between 0 and 100~cm of comet surface within 10 comet years.
Our results also suggest that the non-volatile species would concentrate several meters beneath the surface before it hit the early Earth.
This specific mass of non-volatile species may provide unique chemical condition to the volcanic hot spring pools.
}
   {}


   \maketitle
%

\section{Introduction}

Since the pioneering suggestion by \cite{Oro61}, many researchers discussed the possibility of external delivery of organic materials by comet or meteorites as the supplier of organic materials to the early Earth.
\cite{Ehrenfreund02} strengthened the importance of comet for the sources of organic materials compared to other possible sources on the Earth, such as UV photolysis or the electron discharging in the ancient atmosphere.
If this is the case, an attempt to understand the origins of life must necessarily begin with detailed studies of the formation and evolution of complex organic molecules.
These organic molecules are the products of complex chemistry that most likely starts in molecular clouds and continues within the protoplanetary disk, which is the birthplace of comets and asteroids.
The detections of various kinds of organic materials in the Murchison meteorites, such as sugars and amino acids \citep{Cooper01, Engel82}, support this theory.

If this is the case, the chemical evolution of material should start from the parent cloud of the Solar System.
Recent observations improved our understanding of the chemical compositions in the protoplanetary disk and star-forming regions.
The high sensitivity observation of ALMA achieved the detections of complex organic molecules CH$_3$CN \citep{Oberg15, Loomis18, Bergner18}; CH$_3$OH \citep{Walsh16}; and HCOOH \citep{Favre18}.
Besides, observations toward low-mass star-forming regions are much easier than protoplanetary disks due to high beam-averaged column densities of molecules.
More complex molecules such as glycolaldehyde (HCOCH$_2$OH) \citep{Jorgensen12}, methyl isocyanate (CH$_3$NCO)\citep{Martin-Domenech17}, and CH$_3$Cl \citep{Fayolle17} were reported towards low-mass star-forming regions. 
Glycine (NH$_2$CH$_2$COOH), the simplest and only non-chiral member out of 20 standard amino acids, has gathered the attention of astronomers; however, none of the observations were successful toward any star-forming regions \citep[e.g.,][]{Ceccarelli00}.
The astrochemical modeling of interstellar glycine predicted that CH$_3$NH$_2$ and CH$_2$NH would be potential precursors of glycine \citep{Garrod13, Suzuki18b}.
The detections of CH$_3$NH$_2$ and CH$_2$NH were reported towards various high-mass star-forming regions \citep{Suzuki18a, Ohishi19}.
For glycine itself, \cite{Kuan03} claimed the first detections of glycine towards high-mass star-forming regions, but their detections were refuted by \cite{Snyder05}.

It is thought that the materials in star-forming regions and protoplanetary disks could be incorporated into the comet.
\cite{Walsh14} performed the chemical modeling study for the protoplanetary disk and showed that grain-surface fractional abundances of simple  and complex organic molecules (relative to water ice) for the outer disk are consistent with abundances derived for comets.
\cite{Biver15} also claimed good agreement of the chemical compositions between the cometary coma and warm cores.
The amino acids are also a major target for the in-situ or sample return mission missions of comets.
\cite{Elsila09} claimed the detection of glycine in the Stardust sample with its carbon isotopic ratio $\delta^{13}C=+29\pm6$~‰, suggesting the extraterrestrial origin.
The latest paradigm shift of our knowledge was brought about by the detailed analysis of comet 67P/Churyumov-Gerasimenko (hereafter 67P) by Rosetta mission, enabling us the direct measurement of its chemical composition.
%
%
This mission confirmed all the known chemical species at that time \citep{Roy15}, and new detections of various complex species as well (e.g., CH$_3$NCO, CH$_3$COCH$_3$,  C$_2$H$_5$CHO, and CH$_3$CONH$_2$) through the direct analysis of COmetary Sampling And Composition experiment COSAC \citep{Goesmann15}, though the detection of CH$_3$NCO was disputed by the proceeding study \citep{Altwegg17}.
Recent measurement suggested the existence of ammonium salt in this comet, which could be the substantial reservoir of nitrogen \citep{Altwegg20,Poch20}.
In addition, the volatile glycine and its precursor, CH$_3$NH$_2$, were detected in the coma of comet 67P by the ROSINA (Rosetta Orbiter Spectrometer for Ion and Neutral Analysis) mass spectrometer \citep{Altwegg16}. \cite{Hadraoui19} argued that glycine would originate from sublimating water ice from dust particles that are ejected from the nucleus.

Though this scenario is fascinating, there would be room for discussing the cometary surface's chemistry and physics. 
As \cite{Capaccioni15} reported the very low albedo on the surface of 67P, water rapidly sublimates from the surface, and refractory materials are left behind.
\cite{Capaccioni15} also suggested that refractory organic materials cover the comet surface.
The Deep Impact mission \citep{Mumma05} reported the significant increase of a volatile species, C$_2$H$_6$, after impact to comet 9P/Tempel~1.
The sudden increase of C$_2$H$_6$ implies that the volatile species were lost from the comet surface due to thermal processing, while it is still frozen below the surface of the comet.
This observation also emphasizes the importance of thermal processing to discuss the chemical compositions close to the comet surface.
During the thermal processing by the Sun, volatile species would be easily lost, while non-volatile species would concentrate at the surface, making the non-homogeneous distributions of molecules. 

Comets would have hit the early Earth providing the materials.
\cite{Pierazzo99} performed hydrocode simulations of the comet impact and studied the thermal decomposition rates of amino acids based on the time evolution of the temperature after the impact.
As a result, they showed that the opposite side to the collision point is a good place where amino acids could survive the high-temperature condition during the impact.
If this is the case, it will provide the high-density organic mass without diluting to the volcanic hot spring pools, which is suggested as the candidate place for the origin of life \citep{Damer20}.
As the concentration of non-volatile species gives the unique chemical composition to the volcanic hot spring pools, it is interesting to investigate the chemical distribution in the cometary interiors.
Besides, a previous modeling study by \cite{Keller15} overestimated the water production rate of 67P.
The concentration of the non-volatile species may reduce the evaporation of water and other species from the comet's interior, which may improve the agreement between the modeled and the observed water sublimation rate.

In this paper, we develop a new physical model for the evolution of the cometary interior, including the evaporation processes of water, CH$_3$NH$_2$, and glycine. In Section 2, we describe the details of the Simplified Cometary Concentration Model (SCCM).
Then, in Section 3, we benchmark SCCM with the production rates of water and CH$_3$NH$_2$ and the layer's thickness composed of the non-volatile species. We summarize our work in Section 4.

\begin{figure}
 \begin{tabular}{ll}
\includegraphics[scale=.4]{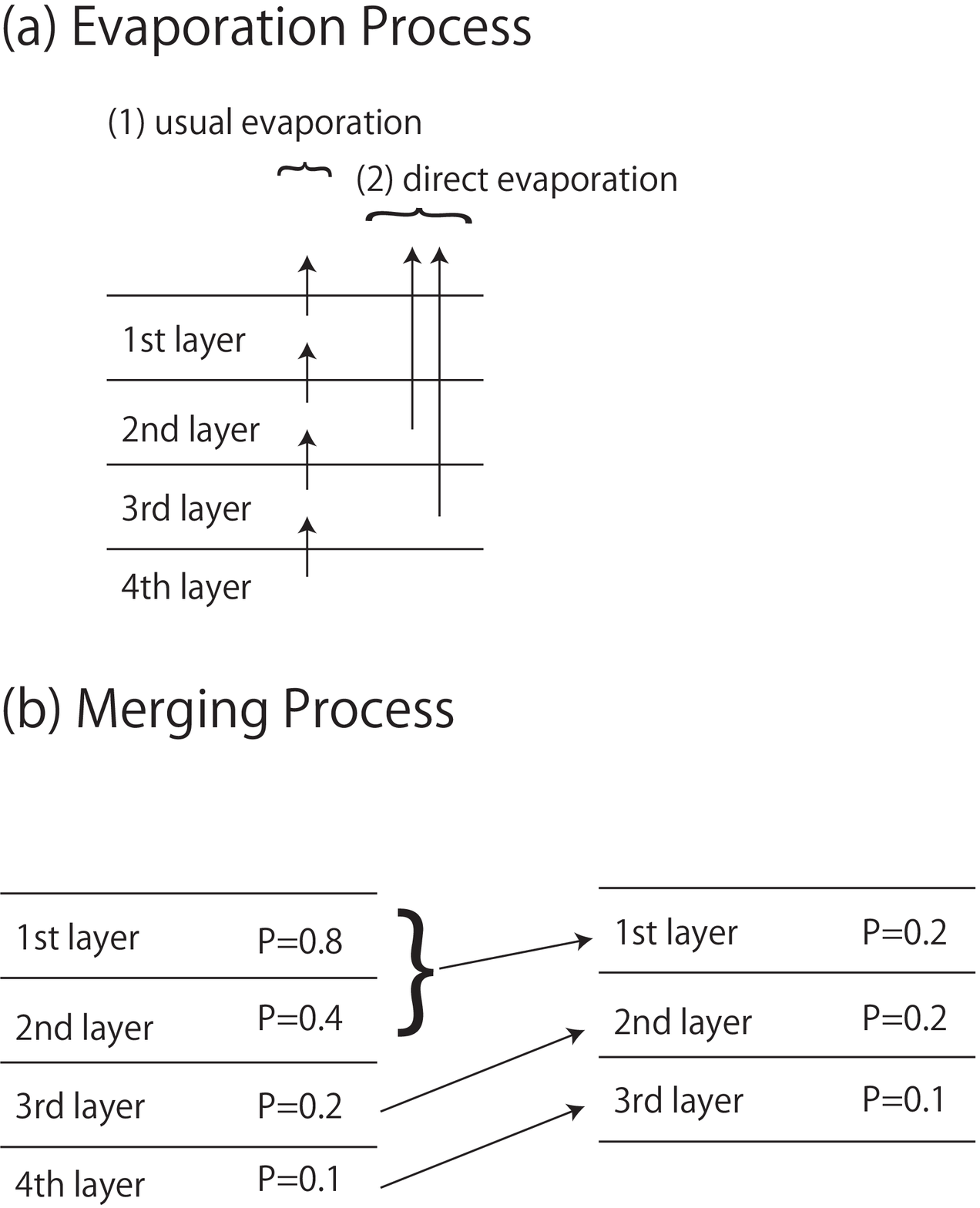}&\\
  \end{tabular}
\caption{
Our SCCM model includes two evaporation mechanisms.
In addition to the usual evaporation process to transport molecules towards upper layer, we consider the direct evaporation process whose efficiency strongly depends on the porosity of upper layers.
When the porosity exceeds 0.7, the layer will merge with the beneath layer.
\label{fig:evaporation_model}
}
\end{figure}

\section{Simplified Cometary Concentration Model (SCCM)}
\subsection{Calculation Method of SCCM}
In the SCCM, we assume a comet's layer structure.
Our layer structure is depicted in Figure~\ref{fig:evaporation_model}.
The thickness of the layer is the parameter that determines the resolution of the vertical structure.
While the smaller size of the layer improves the accuracy of the model, it requires higher computational time. 

\begin{figure*}
 \begin{tabular}{l}
\includegraphics[scale=.6]{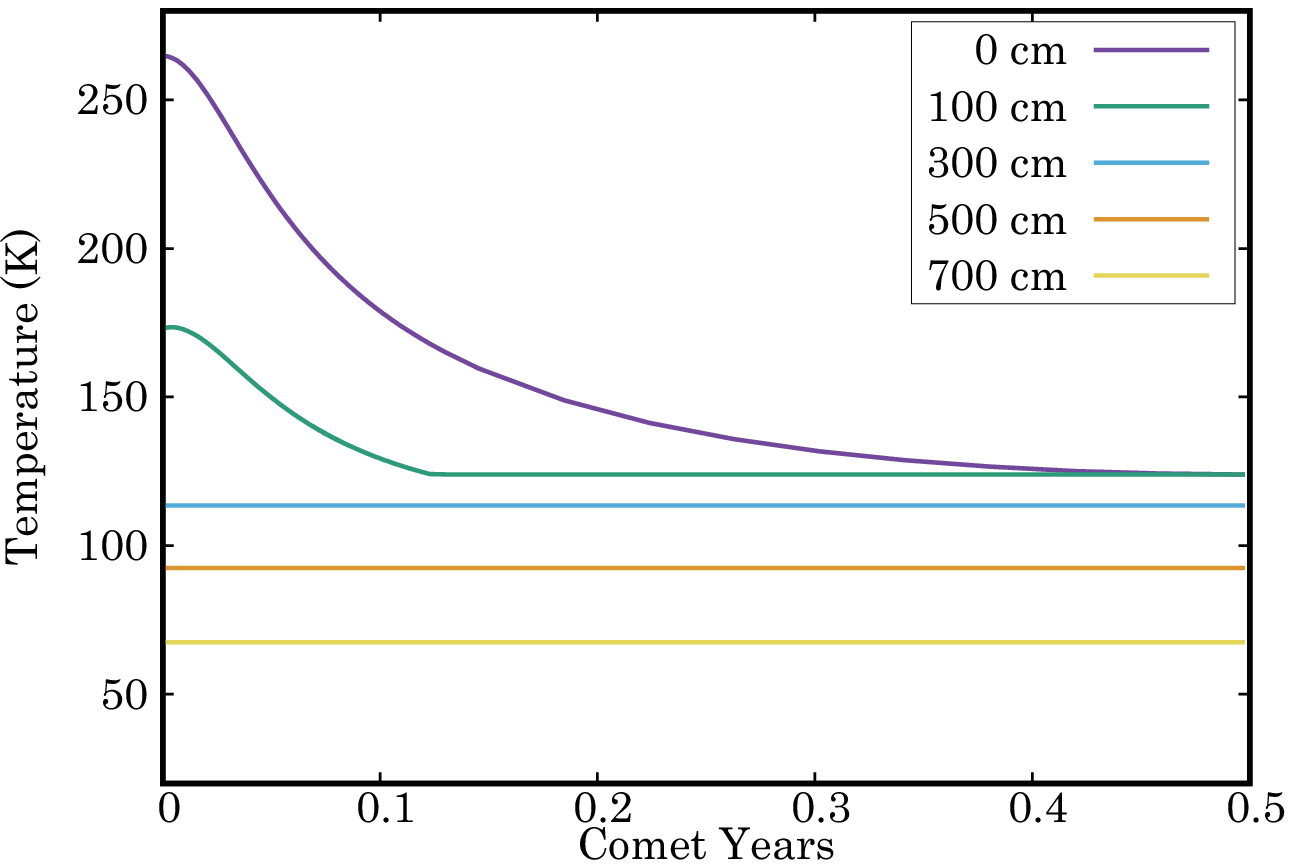}
\includegraphics[scale=.6]{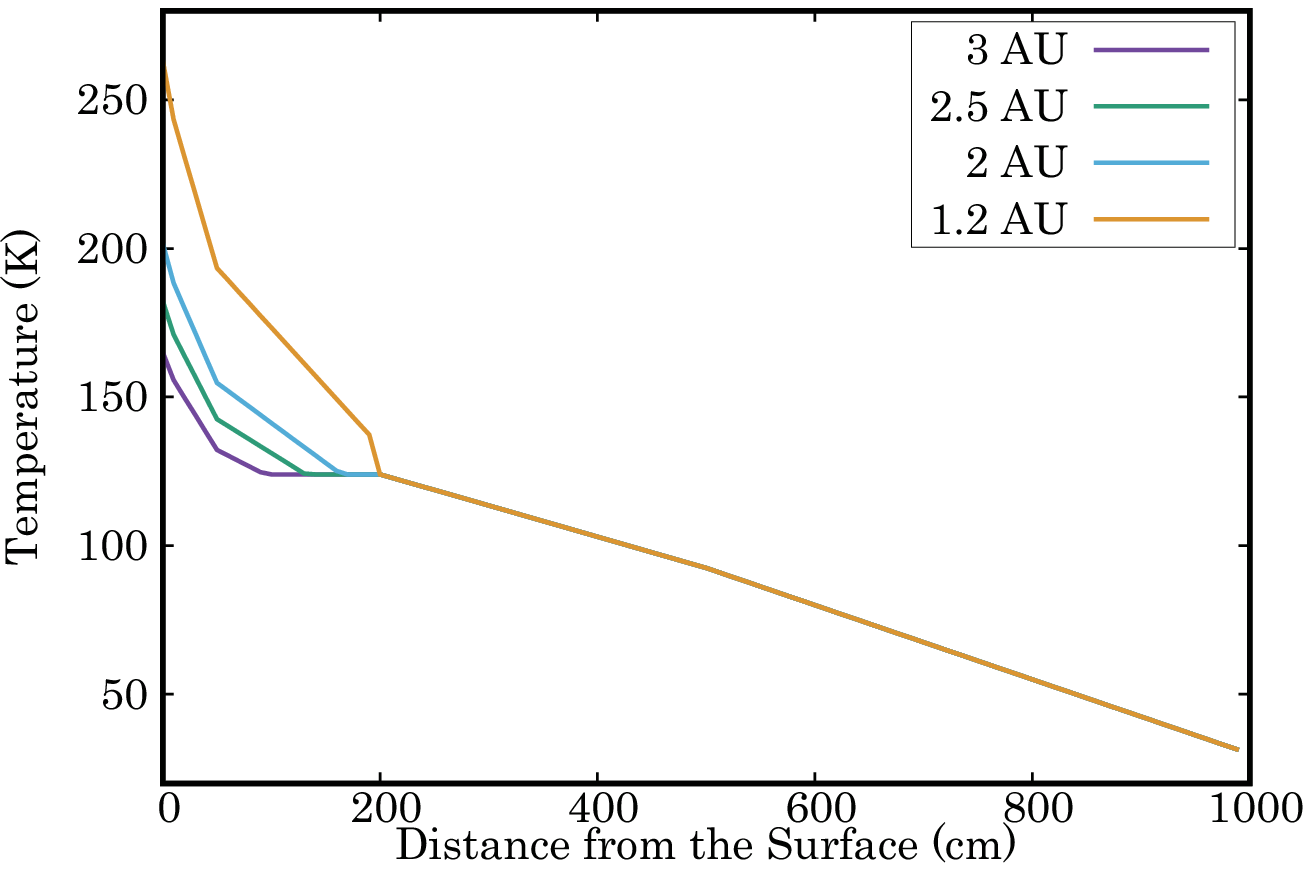}
  \end{tabular}
\caption{
The distribution of temperature inside the comets. 
(Right) The temperature distributions in the different distance from the surface along with the comet years. The time of 0~year is corresponding to the position of perihelion.
(Left) The vertical temperature distribution of comet surface at different distances from the Sun.
\label{fig:temperature}
}
\end{figure*}

In the SCCM model, we calculate the temperature of the n$^{th}$ layer at the time $t$, $T[n,t]$, with the following procedure.
$t$ is the ``comet years'', where one comet year is 6.45~years according to JPL Small-Body Database Browser\footnote{https://ssd.jpl.nasa.gov/sbdb.cgi}, corresponding to one orbital period of the comet 67P.
First, we simulate the orbit of 67P with its orbital parameters to calculate the surface temperature at time $t$, $T[n=1,t]$.
Then, we assume the radiative equilibrium between the irradiation from the Sun and the blackbody radiation of the comet.
The obtained temperature is in good agreement with that of \cite{Guilbert-Lepoutre16} for the surface.
For simplicity, we calculate the surface temperature only from zero to 0.5 comet year and assume the periodic time variation after that.
We note that this symmetric temperature variation is not strictly correct, as the actual peak activity of 67P is observed two weeks after the perihelion \citep{Snodgrass16}.
We use the temperatures of 260, 240, 190, and 120~K for the depths of 0, 10, 50, and 200~cm, respectively, at the perihelion ($t$=0), following the calculation of \cite{Guilbert-Lepoutre16}.
The peak temperature of other depth, $T[n,t=0]$, were linearly interpolated. 
Following \cite{Guilbert-Lepoutre16}, we assume that the temperature is 120~K for all layers at 0.5 comet years, when the comet is the most distant from the Sun. 
Then, for the depth of less than 200 cm, we calculate the temperature of the n$^{th}$ layer at the time $t$ as 
\begin{equation}
T[n,t] = T[n=1,t] \times f[n,t],
\end{equation}
where $f[n,t]=2 \times (1-q[n]) \times t + q[n]$ and $q[n]=T[n,t=0]/T[n=1,t=0]$.
We neglect the time variation of the temperature for the layers deeper than 200~cm.
We fix the temperature of 120 and 30~K at a depth of 200 and 1000~cm, respectively, and use linear interpolation at other depths. 
There is no temperature gradient inside the single layer.
We show our temperature distribution in Figures~\ref{fig:temperature}.
The right figure shows the time dependency of the temperature distribution inside the comets, while the left figure represents the vertical temperature distribution at the different depth from the surface at the different distance from the Sun.

With the temperature, we calculate the evaporation rate from the comet surface with the equation (2) of \cite{Hasegawa92}, which is frequently used in the astrochemical models \citep{Ruaud16}:
\begin{equation}
t[n, j, t]_{\rm evaporation}=\nu_0^{-1}\exp( E[j]/k_{\rm B}T[n, t])\rm{, where,}
\end{equation}
\begin{equation}
\nu_0=(2n_{\rm s}E[j]/\pi^2m)^{1/2}
\end{equation}
The inverse of the evaporation timescale is the evaporation rate $k[n, j, t]$ (i.e.,  $k[n, j, t] = t[n, j, t]_{\rm evaporation}^{-1}$).
$E[j]$ is the binding energy, which is determined by the strength of interaction of species $j$ with water ice, $n_{\rm s}$ is the surface density of the site, which is 1.5$\times10^{15}$~cm$^{-2}$ \citep{Tielens87}, $m$ is the mass of the molecules, and  $k_{\rm B}$ is the Boltzmann constant.
The evaporation rate of species $j$ from $n^{th}$ layer at time $t$, ``$k[n, j, t]$'', is proportional to exp($-E[j]/k_{\rm B}T[n, t]$).
Hence, the binding energy and the temperature are essential to determine $k[n, j, t]$.
We assume that all molecules can evaporate from the surface without being trapped inside.
While the molecules in the deeper layer have to move to upper layers one by one through this usual evaporation process (equation (5)), they can escape directly from the comet from any layer via a direct evaporation process as described below (equation (6)).
%
%
The non-volatile species should be mainly minerals and insoluble organic matter \citep{Fray17}.
The binding energies for H$_2$O, CH$_3$NH$_2$ are 5700 and 6500~K, respectively, based on the theoretical prediction by \cite{Wakelam17}.
We assume that thermal evaporation is negligible for non-volatile species.
For glycine, the binding energy would be 13000~K \citep{Tzvetkov04,Suzuki18b}, and the thermal evaporation is not dominant in the orbit of 67P.
Therefore, we treat glycine as a part of non-volatile species.

In SCCM modeling, we define the porosity $P[n,t]$ as
\begin{equation}
P[n,t]=\displaystyle\frac{V_{\rm Max}-V_{\rm Real}[n,t]}{V_{\rm Max}},
\end{equation}
where $V_{\rm Real}[n, t] = \Sigma_{j} V_{\rm Real}[n, j, t]$ represents the sum of the volumes occupied by materials of all species $j$ on $n^{th}$ layer and at $t$, while $V_{\rm Max}$ denotes maximum volume in a certain layer.
From this definition, it is apparent that the term $V_{\rm Real}[n, t]=(1-P[n, t])\times V_{\rm Max}$ denotes the real volume occupied by any material in a layer.
%
%
Here we introduce the ratio of certain species ``$j$" in n$^{th}$ layer at $t$, $r[n,j,t]$ as $r[n,j,t]=V_{Real}[n,j,t]/V_{Real}[n,t]$.
The sum of the ratios of all molecules, $\sum_j r[n,j,t]$, should always be unity.

With the term of $V_{\rm Real}[n,j,t] = V_{\rm Real}[n,t] \times r[n,j,t] = (1-P[n,t]) \times V_{\rm Max} \times r[n,j,t]$, the evaporated volume amount of species $j$ from the first layer ($n=1$) during the time step $\Delta t$, $V_{\rm evap}[1, j, t]$, is given by 
\begin{equation}
\frac{\Delta V_{\rm evap}[1, j, t]}{\Delta t}=k[1, j, t] \times V_{\rm Real}[1, j, t].
\end{equation}
Since the porosity $P[1,t]$ represents the percentage of cavity, $(1-P[1,t])\times V_{\rm Max}$ represent the actual amount of material in the 1$^{st}$ layer.
The volume amount of material transferred from $n+1^{th}$ layer to $n^{th}$ layer during the time step $\Delta t$ is calculated as 

\begin{equation}
\frac{\Delta V_{\rm evap}[n+1, j, t]}{\Delta t}=P[n,t] \times k[n+1,j, t] \times V_{\rm Real}[n+1,j,t] \quad(n\geqq1).
\end{equation}
Although the default time step $\Delta t$ is 1000~seconds, the time step is half that value when (1) the evaporation rate of at least one species $j$, $\Delta t \times k[n , j, t]$, is less than 1, and (2) the molecular ratio of species $j$, $r[n,j,t]$, is more than 1$\%$.
The first criteria will enable us to perform accurate calculation of volatile species, especially H$_2$O.
When $\Delta t \times k[n, j, t]$ exceeds 1 and $r[n,j,t]$ is less than 1$\%$, the term of $\Delta t \times k[n, j, t] = 1$, corresponding to the situation that all material is transferred to the upper layer during $\Delta t$.
Similar to the equation for the evaporation process from the first layer, $k[n+1, j, t] \times V_{\rm Real}[n+1, j, t]$ denotes the total volume amount of thermal evaporation for species $j$.
However, this equation represents the blocking of material transfer to the upper layer by multiplying the porosity of the upper layer, $P[n, t]$ (Figure~\ref{fig:evaporation_model}).
%

We assume that a part of the material can evaporate directly from any layer if there is enough porosity in the upper layer (Figure~\ref{fig:evaporation_model}).
The volume amount of direct evaporation for species $j$ from the n$^{th}$ layer is presented as 
\begin{equation}
\Delta V_{\rm direct}[n+1, j, t]=(\prod_{k=1}^{k=n} P[k,t]) \times \Delta V_{\rm evap}[n+1, j, t].
\end{equation}
By multiplying the term of $\prod_{k=1}^{k=n} P[k, t]$, this equation permits to evaporate materials directly from the inner layer when the porosity of the upper layers are high enough.

The evaporation processes (5) to (7) determine the volume amount of lost material from the layer.
The differentiation of porosity and molecular ratios can be calculated as the result of the above processes.
The differentiation of $V_{\rm Real}$ during the time step $\Delta t$ can be calculated as
\begin{equation}
\begin{split}
\Delta V_{\rm Real}[n, j, t] = -\Delta t \times (\Delta V_{\rm evap}[n, j, t]/\Delta t \\-\Delta M_{\rm evap}[n+1, j, t]/\Delta t + \Delta V_{\rm direct}[n, j, t]/\Delta t).
\end{split}
\end{equation}
Then, $V_{\rm Real}$ after the time step $\Delta t$ is presented as 
\begin{equation}
V_{\rm Real}[n, j, t+\Delta t]=V_{\rm Real}[n, j, t]+\Delta V_{\rm Real}[n, j, t].
\end{equation}
Then, recalling the definition of the porosity, new porosity $P[n, t+\Delta t]$ is calculated as
\begin{equation}
P[n, t+\Delta t] = \frac{V_{\rm Max}-\Sigma_j V_{\rm Real}[n, j, t+\Delta t]}{V_{\rm Max}}.
\end{equation}
Since $V_{\rm Real}[n,j,t]=(1-P[n,t]) \times r[n,j,t] \times V_{\rm Max}$, this equation is independent of $V_{\rm Max}$.
The time evolution of molecular ratios is calculated as 
\begin{equation}
\begin{split}
r[n,j,t+\Delta t]=V_{\rm Real}[n,j,t + \Delta t]/V_{\rm Real}[n,t+\Delta t]\\={V_{\rm Real}[n,j,t]+\Delta V_{\rm Real}[n,j,t]}/(1-P[n,t+\Delta t]) \times V_{\rm Max}.
\end{split}
\end{equation}

Finally, we consider merging layers when the porosity gets larger than 0.7 after the above calculations.
This value of 0.7 is determined considering the observed porosity of 67P \citep{Patzold16}. 
In addition, the sum of the actual volume amount of layers, $V_{\rm Real(before)}[n, t]+V_{\rm Real(before)}[n+1, t]$, must be smaller than $V_{\rm Max}$ before the merging process so that the volume amount of new layer does not exceed the limitation after merging.
Then, using the definition of porosity, the merging process is considered when $P[n, t] + P[n+1, t] >1$ is achieved.
With these restrictions, new molecular ratios and the porosity for the $n^{th}$ layer can be calculated from the definition of porosity.
We fix $V_{\rm Max}$ during merging but the real volume amount of material is regarded as the sum of the original two layers.
\begin{equation}
\begin{split}
P_{\rm new}[n, t] = \displaystyle\frac{V_{\rm Max}-(V_{\rm Real(before)}[n, t]+V_{\rm Real(before)}[n+1, t])}{V_{\rm Max}}\\=(P_{\rm before}[n, t] + P_{\rm before}[n+1, t]) -1.
\end{split}
\end{equation}
Then, new molecular ratios can be calculated as
\begin{equation}
\begin{split}
r_{\rm new}[n, j, t]\\
= \displaystyle\frac{(1 - P_{\rm before}[n, t])r[n, j, t] + (1 - P_{\rm before}[n+1, t])r[n+1, j, t]}{1-P_{\rm new}[n, t]}.
\end{split}
\end{equation}
After the merging process, we reassign the layer number to compensate the merged layer.
In this process, the old $n+2^{th}$ layer is regarded as new $n+1^{th}$ layer by applying $r_{\rm new}[n+1, j, t] = r_{\rm before}[n+2, j, t]$ and $P_{\rm new}[n+1,t] = P_{\rm before}[n+2, t]$.
The conservation of the total volume amount of material is kept with the above equations.

The timescale of SCCM is also essential since non-volatile species would efficiently concentrate after the long timescale of the heating by the Sun.
\cite{Ip16} calculated the history of orbital evolution of 67P due to the gravitational interactions between giant planets.
They suggested that the perihelion of comet 67P changed suddenly in 1959 from 2.7 to 1.3~AU.
Before 1959, the perihelion was between 2 and 3~au for 300 years, but it would have the perihelion of $\sim$1.2~au between about 300 about 900~years ago.
Therefore, this comet would have experienced strong heat processing by the solar flux for several hundred years.
To simulate such a long thermal processing timescale, we calculated the time evolution of the porosity and the molecular ratios during 100~comet years.
In the discussion, we will also discuss the case where the comet 67P did not experience the heating before 1959.

Finally, our model is a very simplified one in many points.
Although the photo-dissociation process may destroy the molecules, we do not include it in the SCCM modeling.
According to the experimental result, molecules would be protected from solar UV photons since they can penetrate only thousands of nanometers \citep{Barnett12}.
However, solar and cosmic particles may affect our results by destroying molecules.
In addition, we neglect the sideways and inward diffusion.
While the diffusion sideways will be canceled by the diffusion of opposite direction, the inward diffusion may delay the concentration process.
As we will see in the next section, the concentration process mainly gets completed within 10 comet years.
Therefore, the delay would not be so important if we assume the longer timescale.
Our model does not consider the effect of the rotation of 67P as well.

\section{SCCM Results}
\subsection{Parameter Dependency}
We performed SCCM simulations under the six different sets of chemical ratios, sizes of layer, and initial porosity.
These parameters are summarized in Table~\ref{table:parameter}.
We include H$_2$O and CH$_3$NH$_2$, and non-volatile species.
As water is the dominant component of cometary gas, it is used to simulate the concentration mechanism of non-volatile species.
In addition, we include CH$_3$NH$_2$ as representative of less-volatile species, whose binding energy is higher than water, to test its behavior during the calculation.
We also note that CH$_3$NH$_2$ is known as a precursor of glycine.
Though the list of included species is limited, they will provide us enough information to discuss the concentration mechanism.

\begin{table}
\tiny
\caption{The Parameters for SCCM Modeling}
\label{table:parameter}
\centering  
\begin{tabular}{c|c|c|c}
\hline
\hline
Model & Initial ratios of non-volatile species &  \shortstack {Layer Size \\ (cm)} & Initial Porosity\\
\hline
Set~1&0.5&10&0\\
Set~2&0.5&10&0.2\\
Set~3&0.5&5&0\\
Set~4&0.5&20&0\\
Set~5&0.1&10&0\\
Set~6&0.9&10&0\\
\hline
\hline
\end{tabular}
\tablefoot{
The parameters for SCCM modeling are summarized.
We show the initial volume occupied by non-volatile species, layer thickness, and porosity.
}
\end{table}

We set the initial porosity of zero for Set~1.
Though the porosity of zero is extremely lower than the observation of 67P, we use this porosity to assume an ideal comet, which is free from the heat processing, for the initial condition.
We set the volume ratio of non-volatile species to be 50$\%$.
Other materials are mainly water, but CH$_3$NH$_2$ is also included in our model to see the behavior of molecules with binding energy higher than water.
Considering the simulated abundance ratio of CH$_3$NH$_2$/H$_2$O on interstellar grain, we set the ratio of CH$_3$NH$_2$ to be 0.1$\%$ \citep{Suzuki18b}.
All layers have the same initial molecular ratios regardless of the depth.

\begin{figure*}
 \begin{tabular}{ll}
\includegraphics[scale=.65]{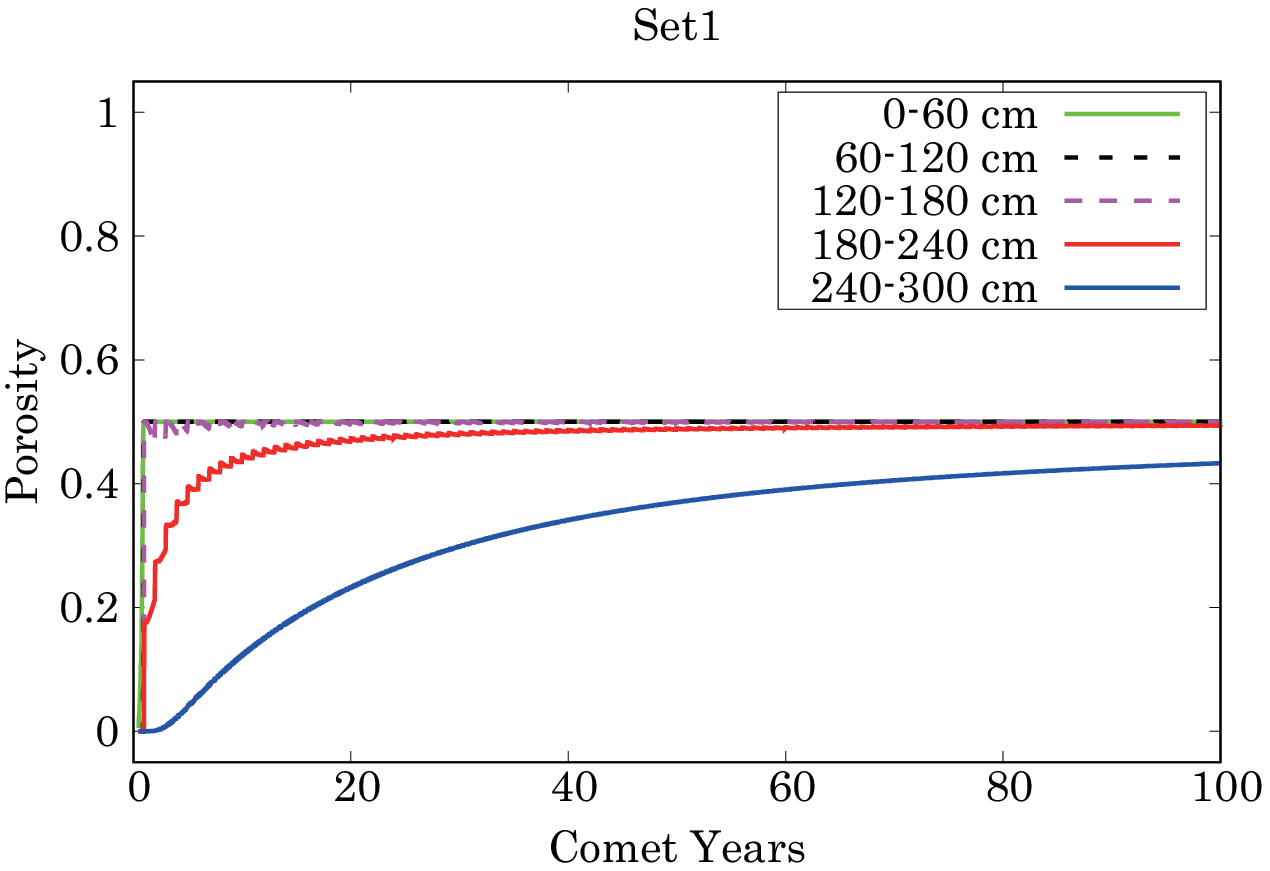}&
\includegraphics[scale=.65]{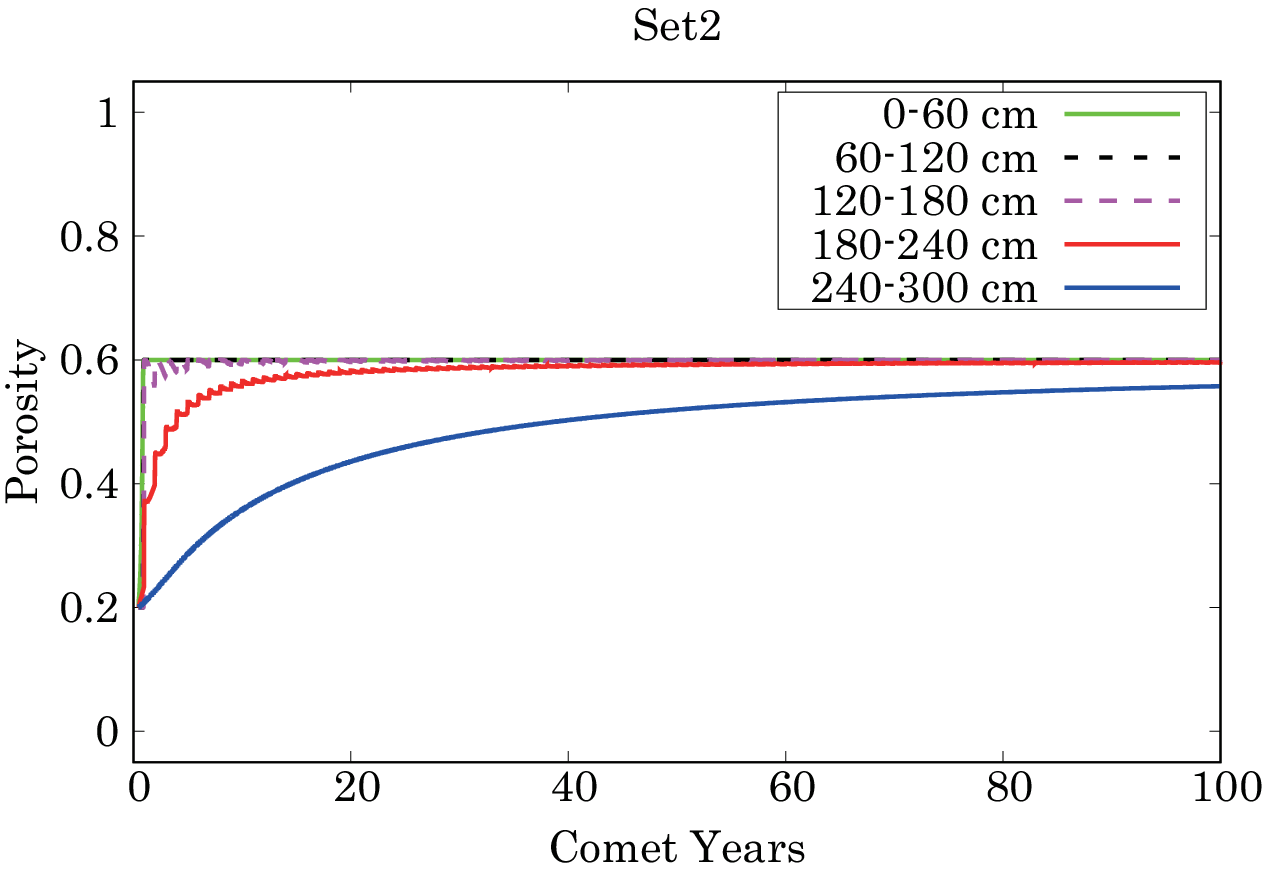}\\
\includegraphics[scale=.65]{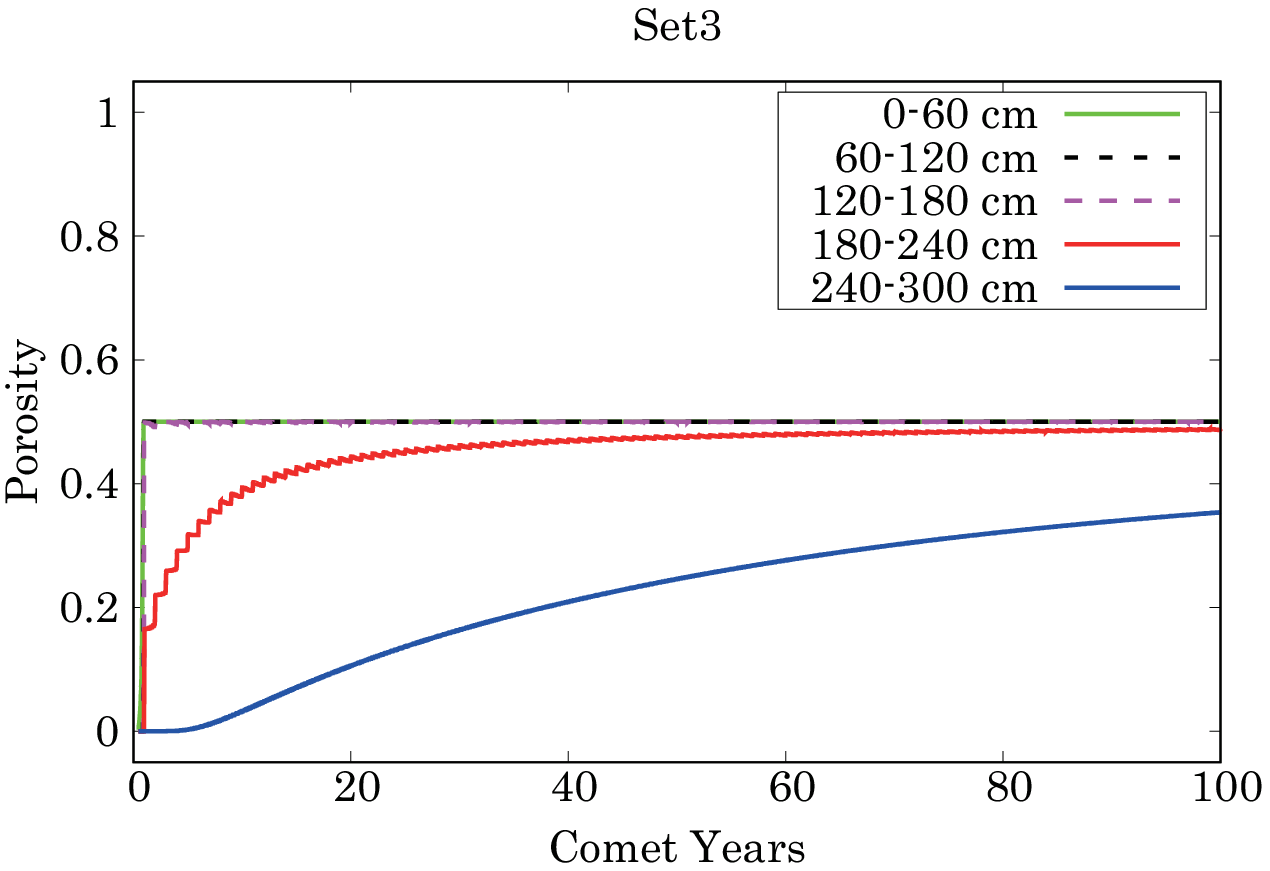}&
\includegraphics[scale=.65]{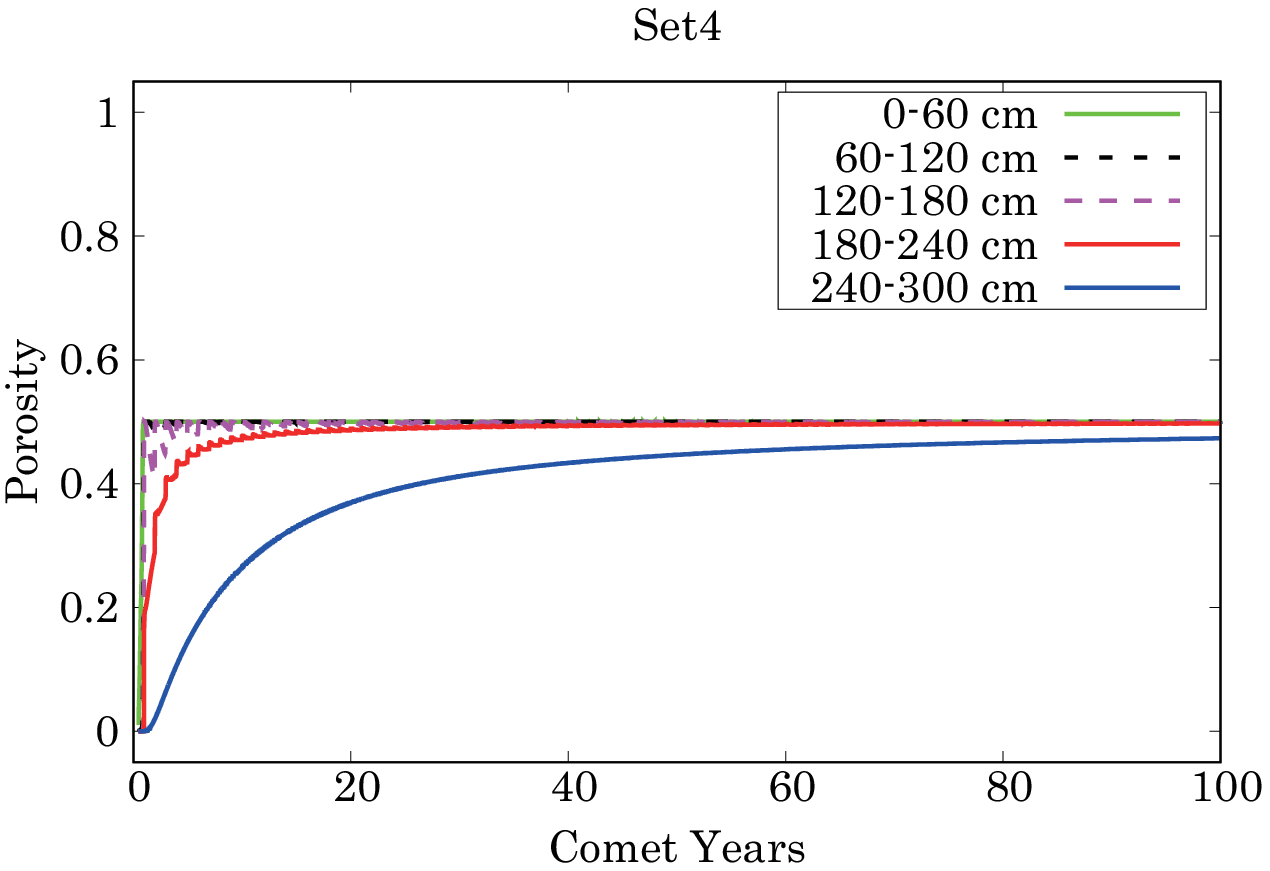}\\
\includegraphics[scale=.65]{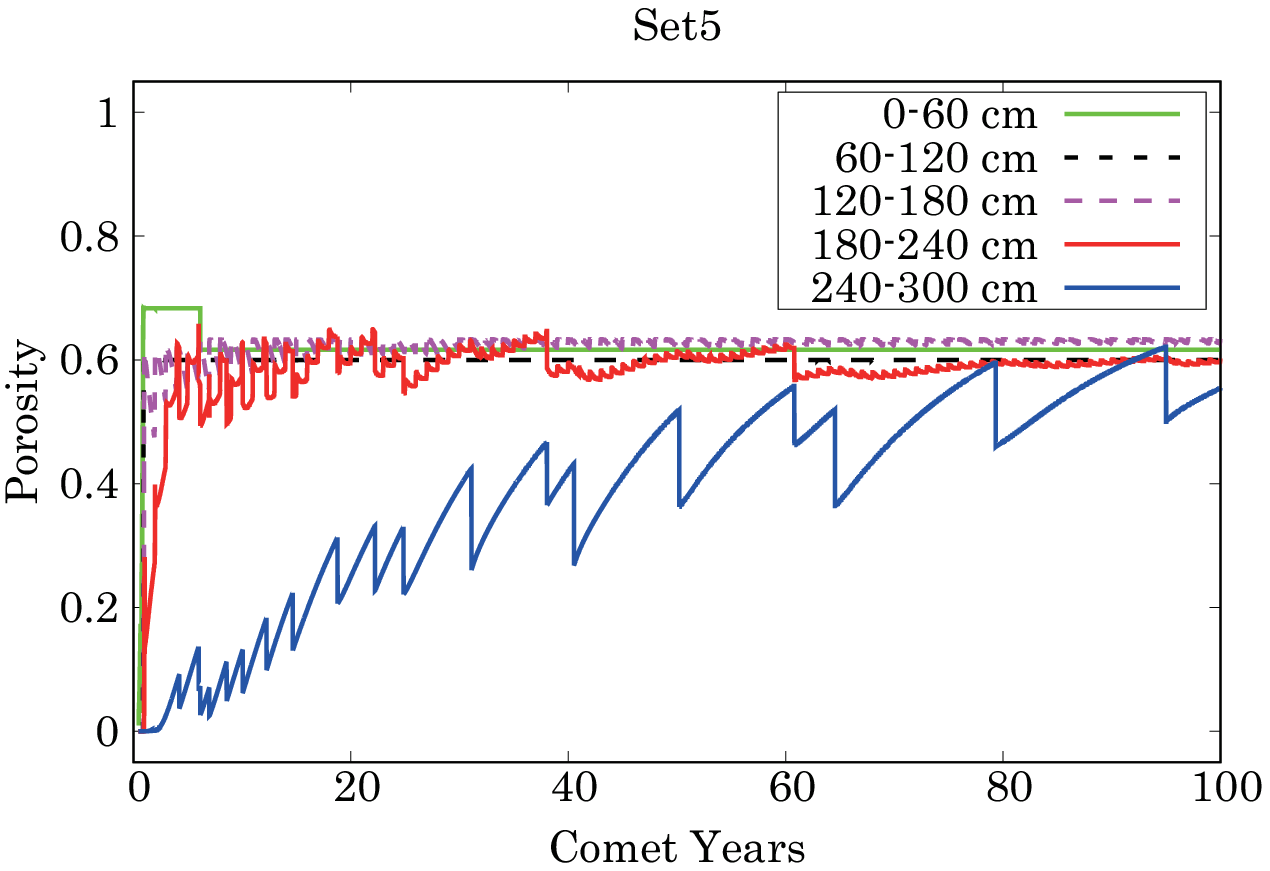}&
\includegraphics[scale=.65]{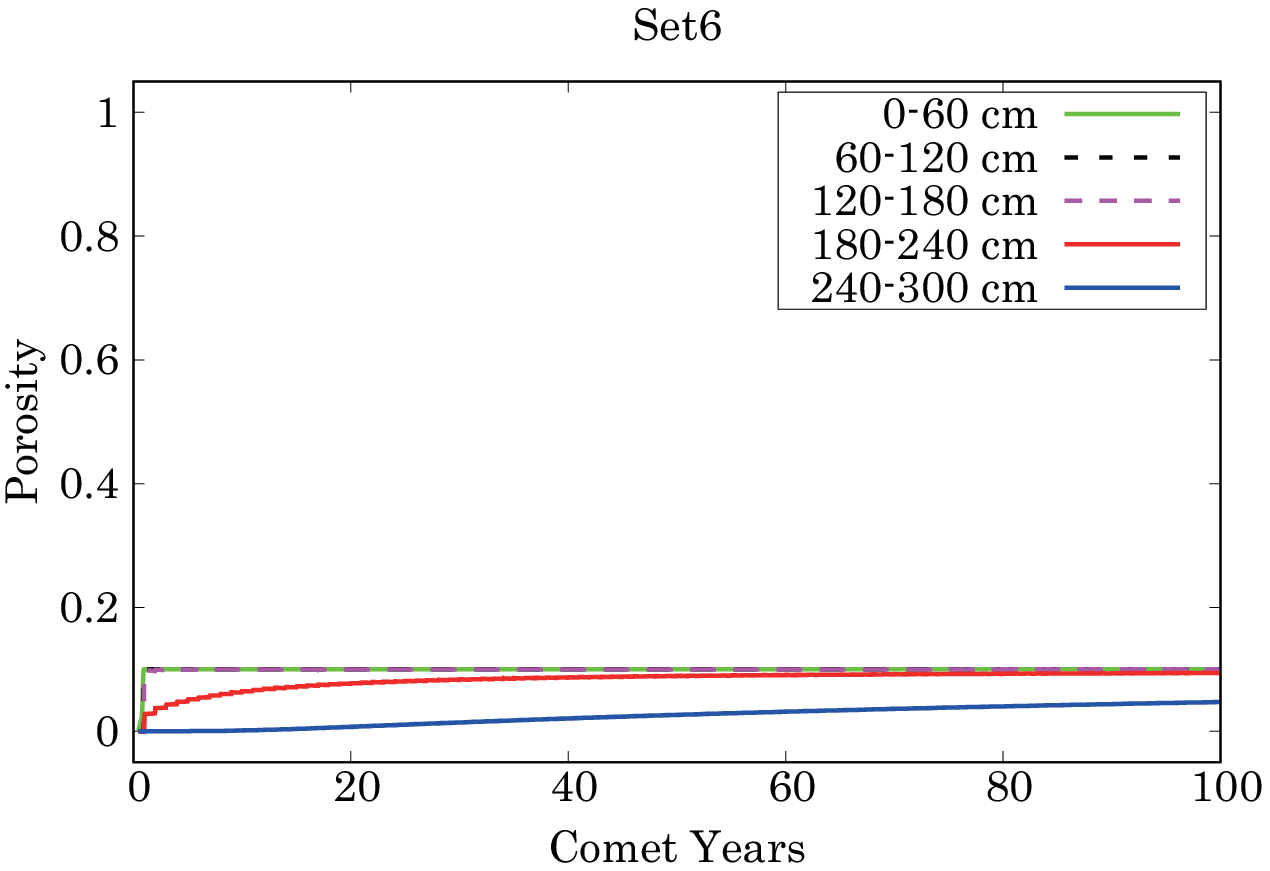}\\
  \end{tabular}
\caption{
The evolution of averaged porosity for the different depths.
The lines represent the depth of 0-60, 60-120, 120-180, 180-240, and 240-300~cm with different colors.
\label{fig:porosity}
}
\end{figure*}

\begin{figure*}
 \begin{tabular}{ll}
\includegraphics[scale=.65]{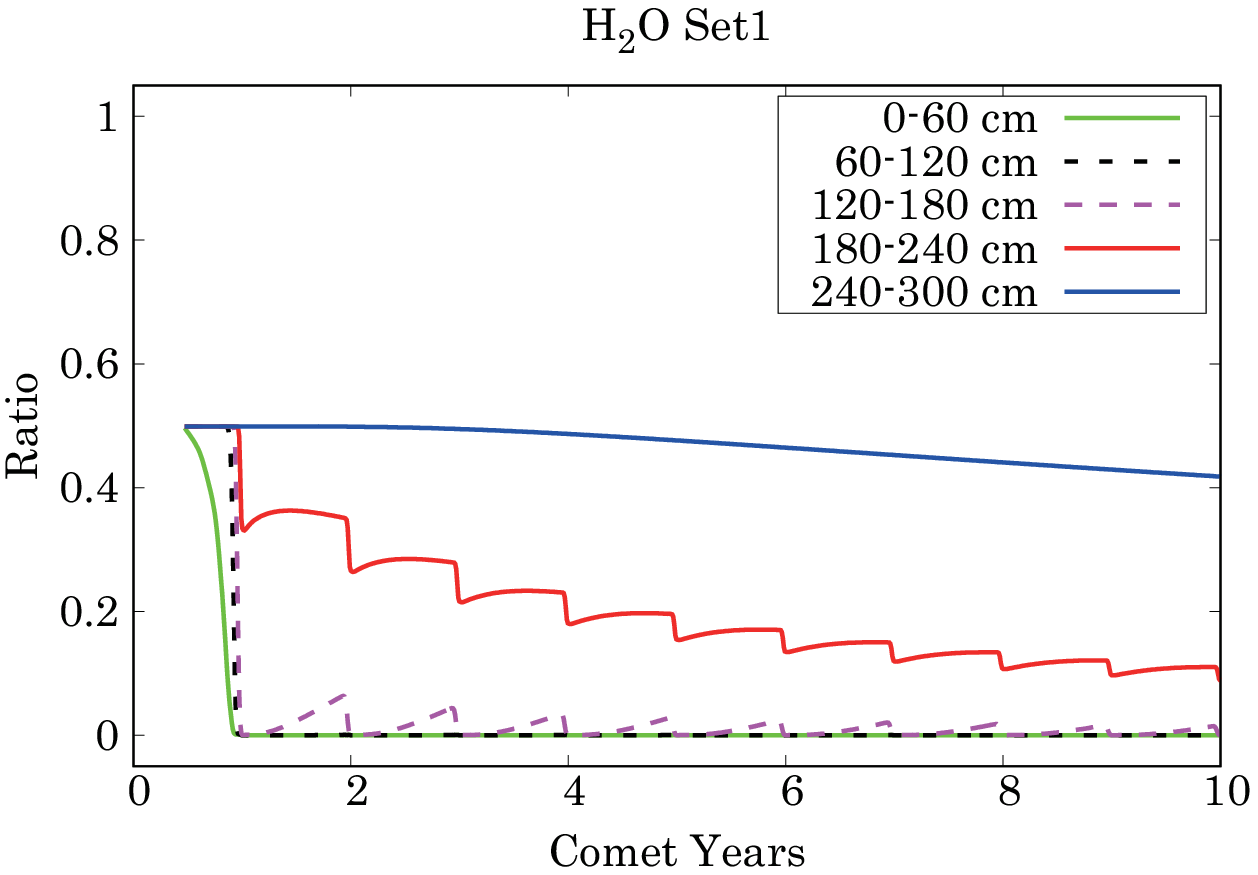}&
\hspace*{0.5cm}
\includegraphics[scale=.65]{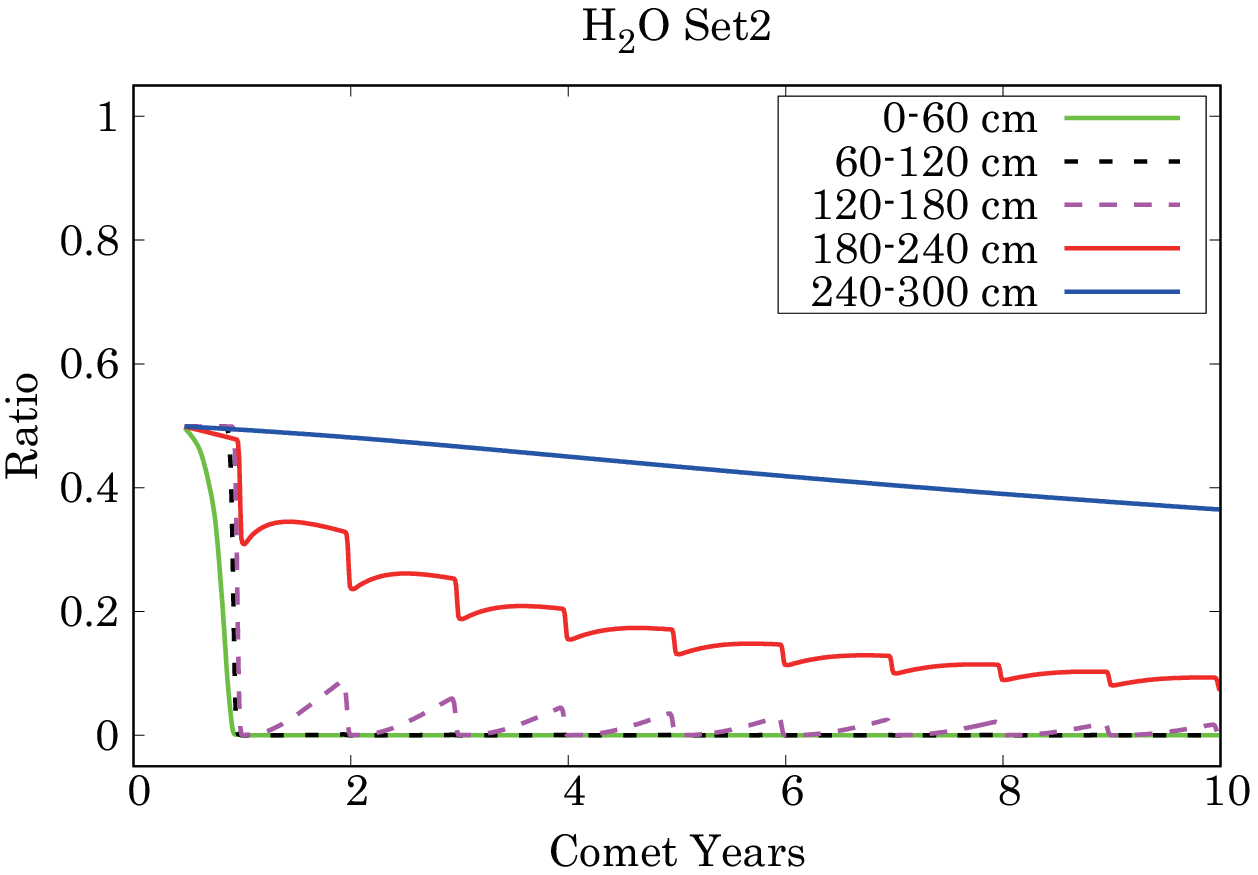}\\
\includegraphics[scale=.65]{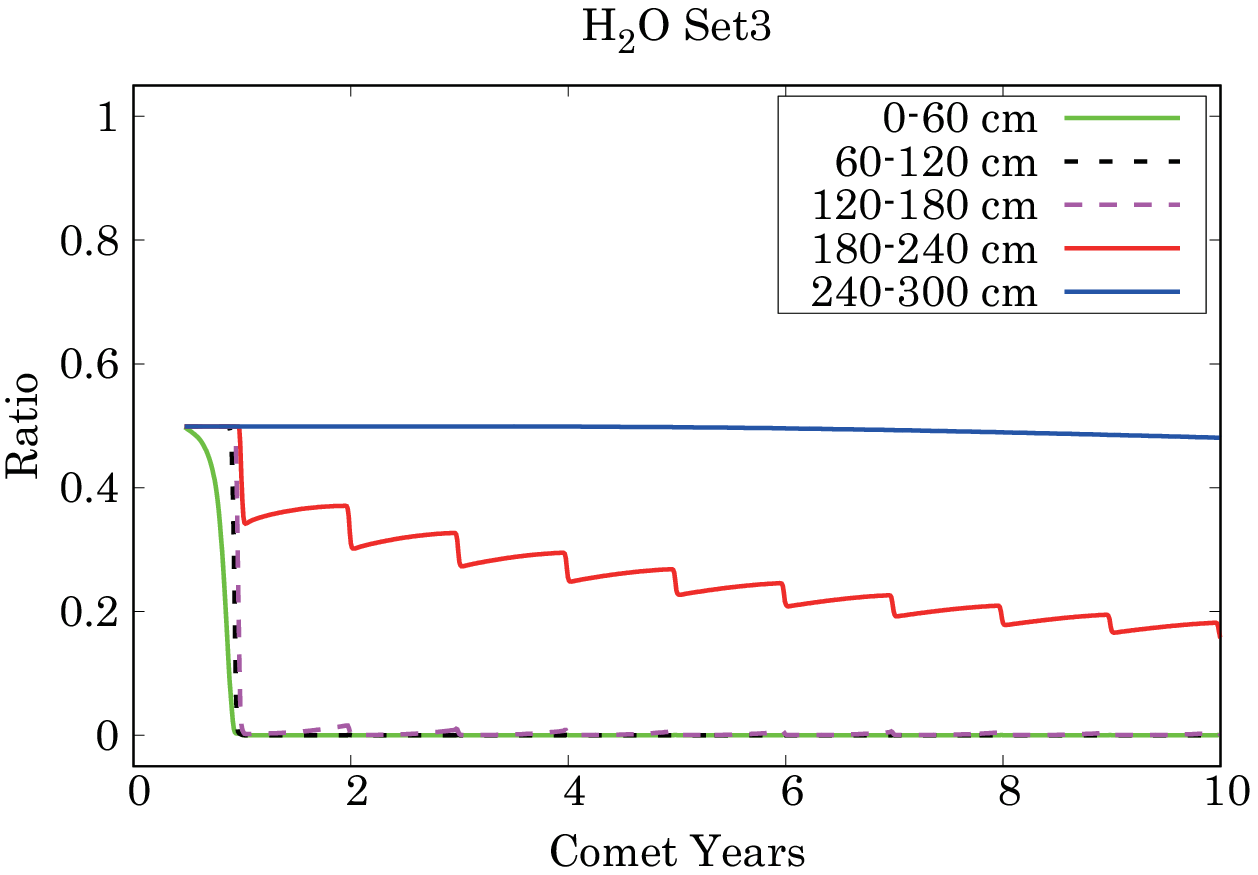}&
\includegraphics[scale=.65]{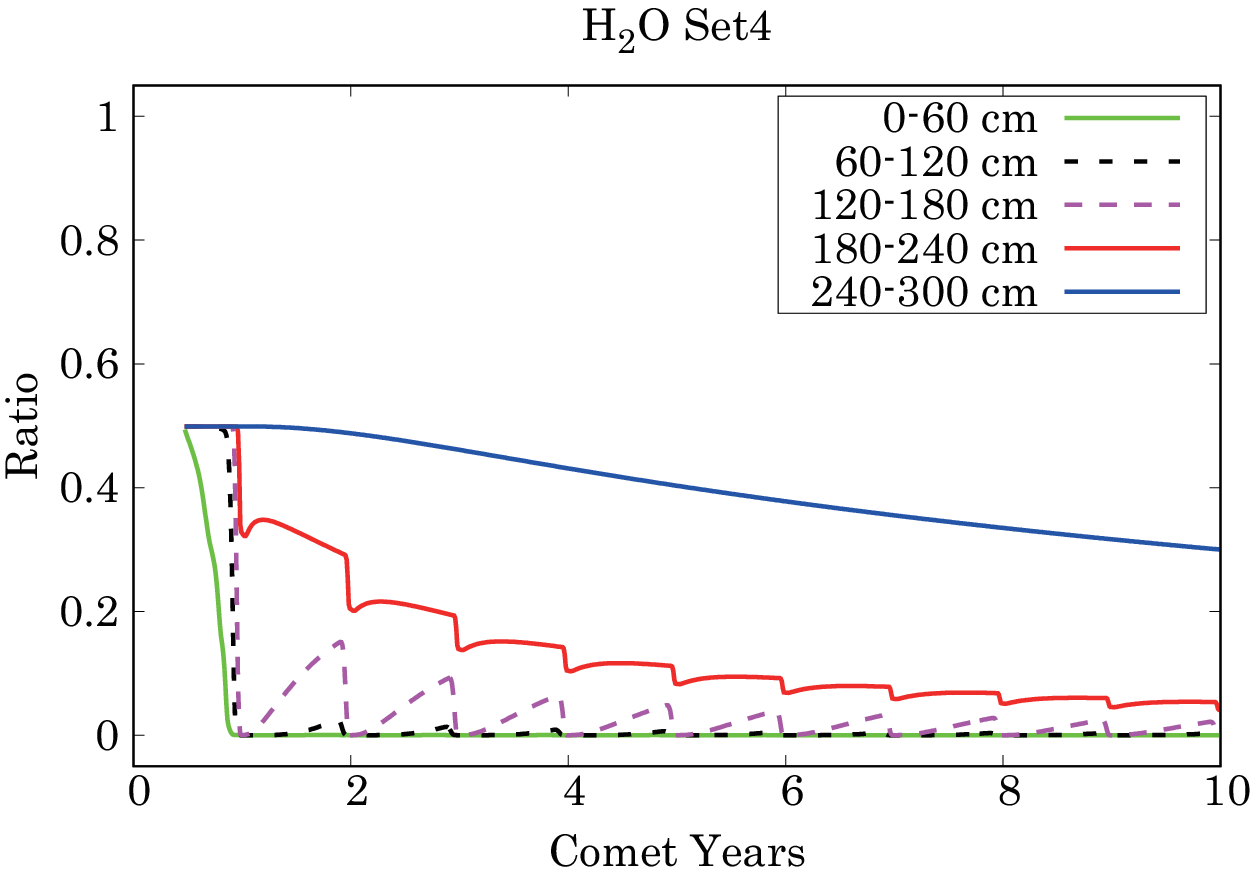}\\
\includegraphics[scale=.65]{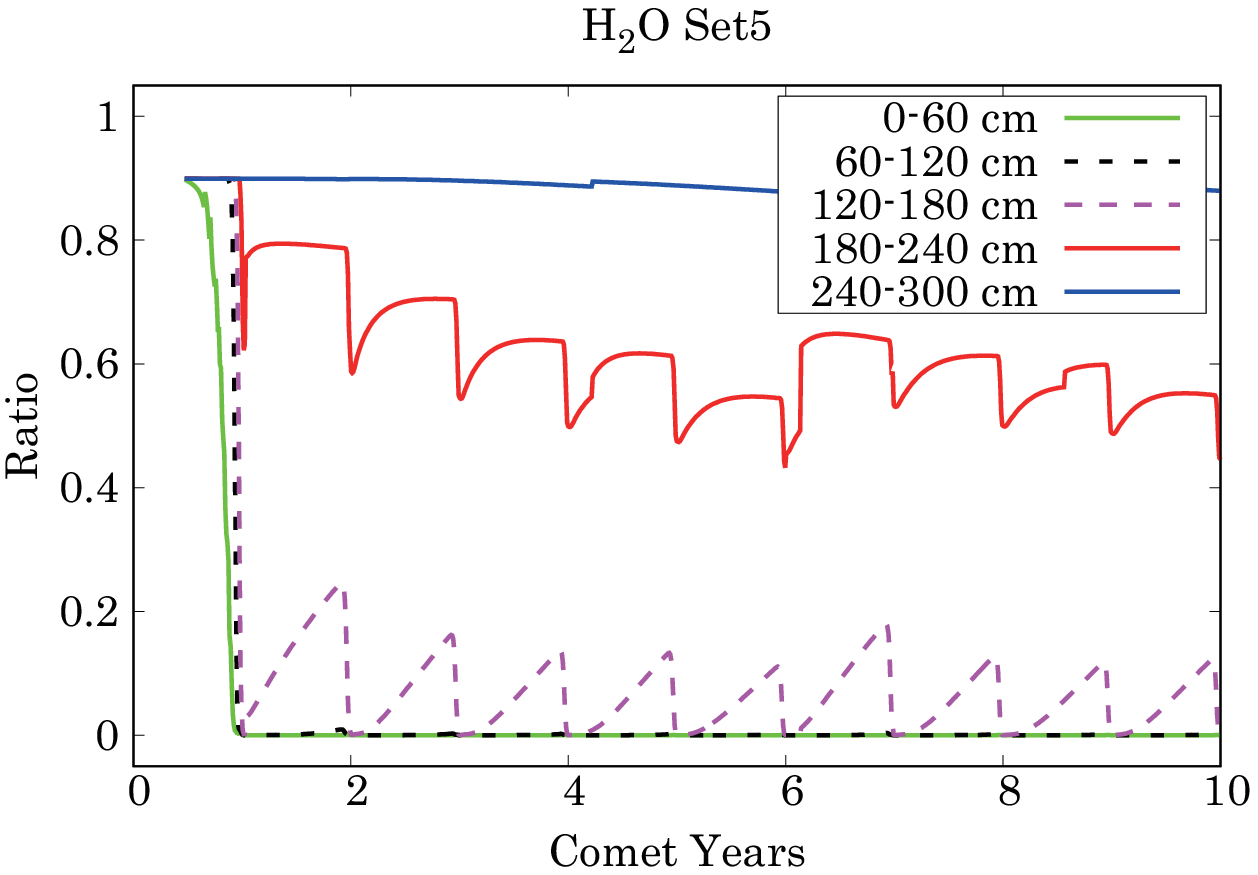}&
\includegraphics[scale=.65]{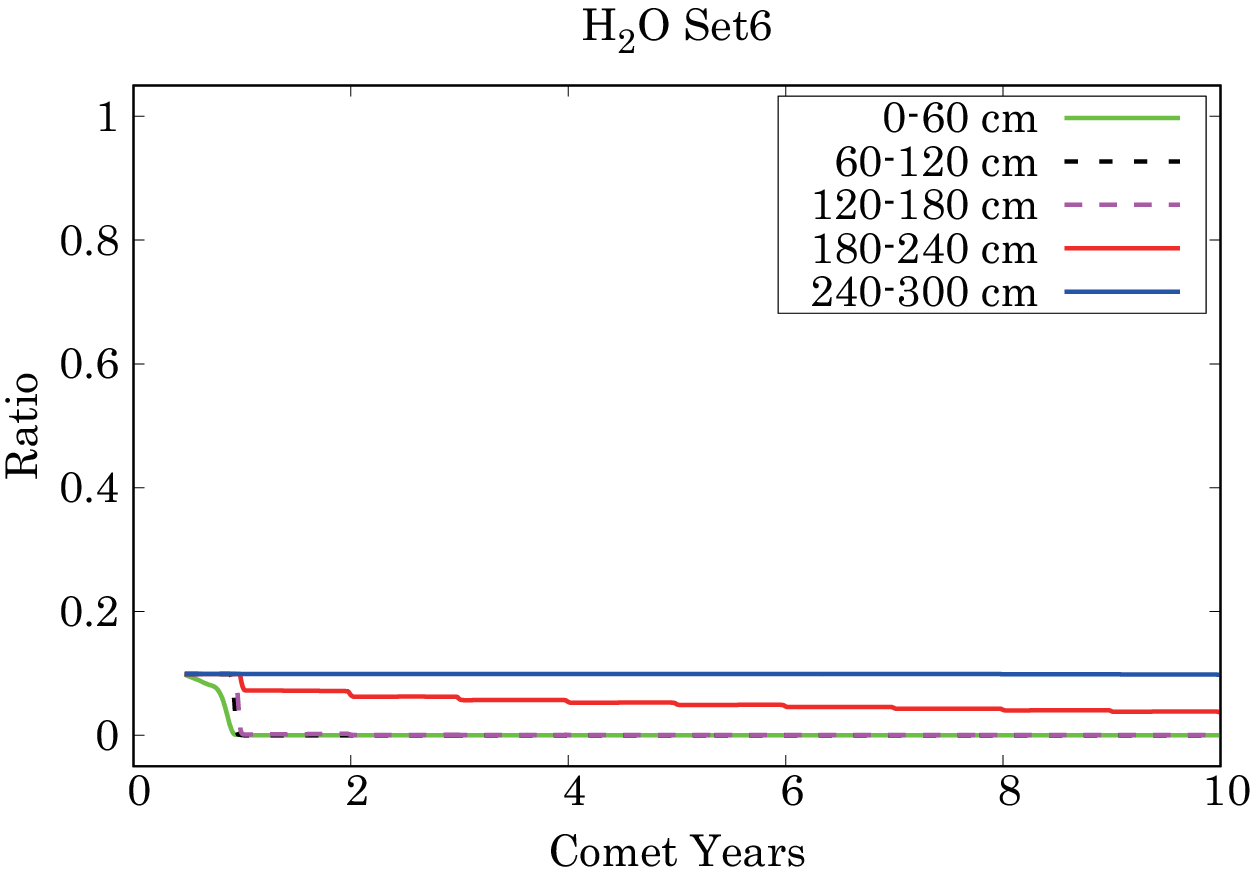}\\
  \end{tabular}
\caption{
With six sets, the time evolution of H$_2$O ratio compared to all volume occupied by materials are shown. 
The sum of the ratios of water, CH$_3$NH$_2$, and non-volatile species is unity.} 
The different colors represents the depths of 0-60, 60-120, 120-180, 180-240, and 240-300~cm.
The horizontal axis represents the time in comet year, with one unit being the one orbital motion around the Sun.
\label{fig:H2O}
\end{figure*}

\begin{figure*}
 \begin{tabular}{ll}
\includegraphics[scale=.65]{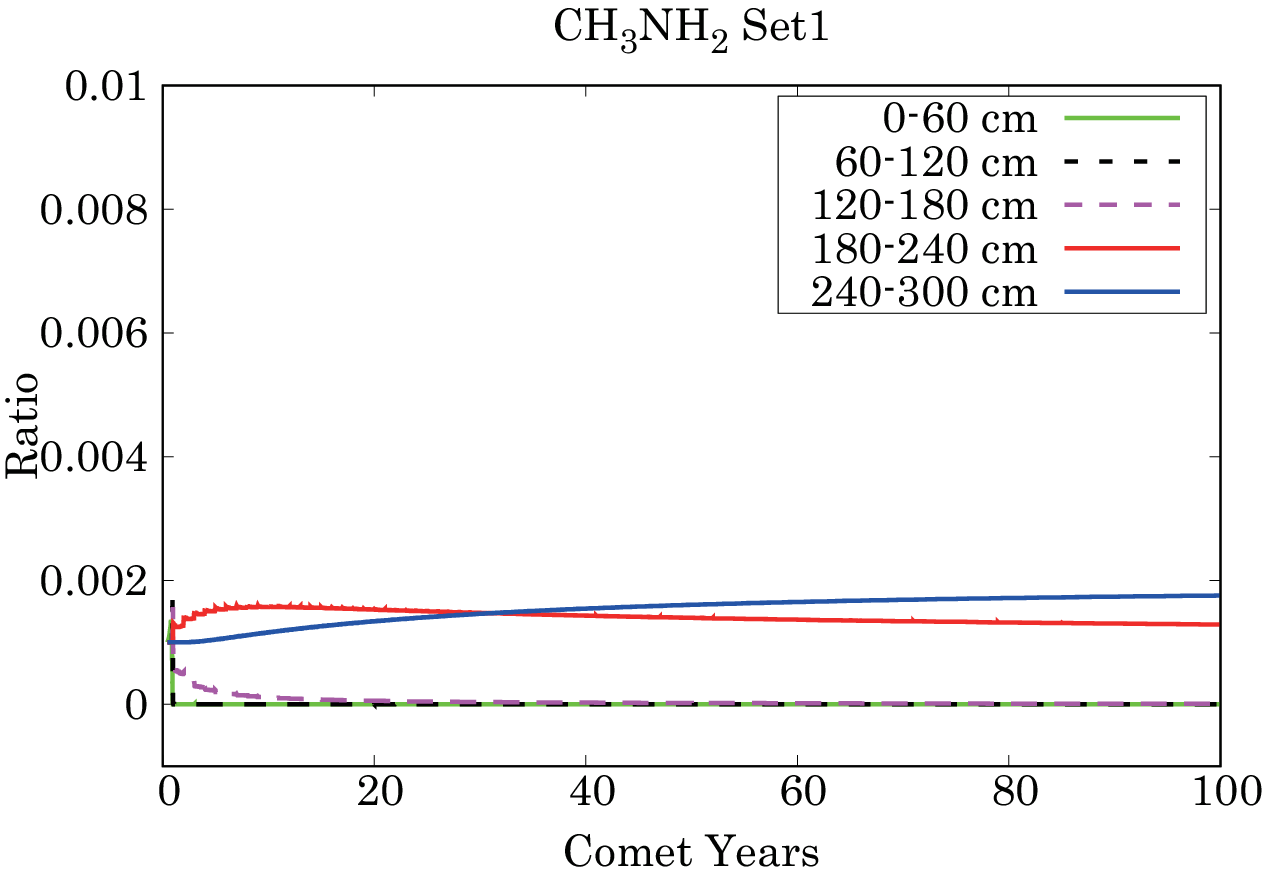}&
\includegraphics[scale=.65]{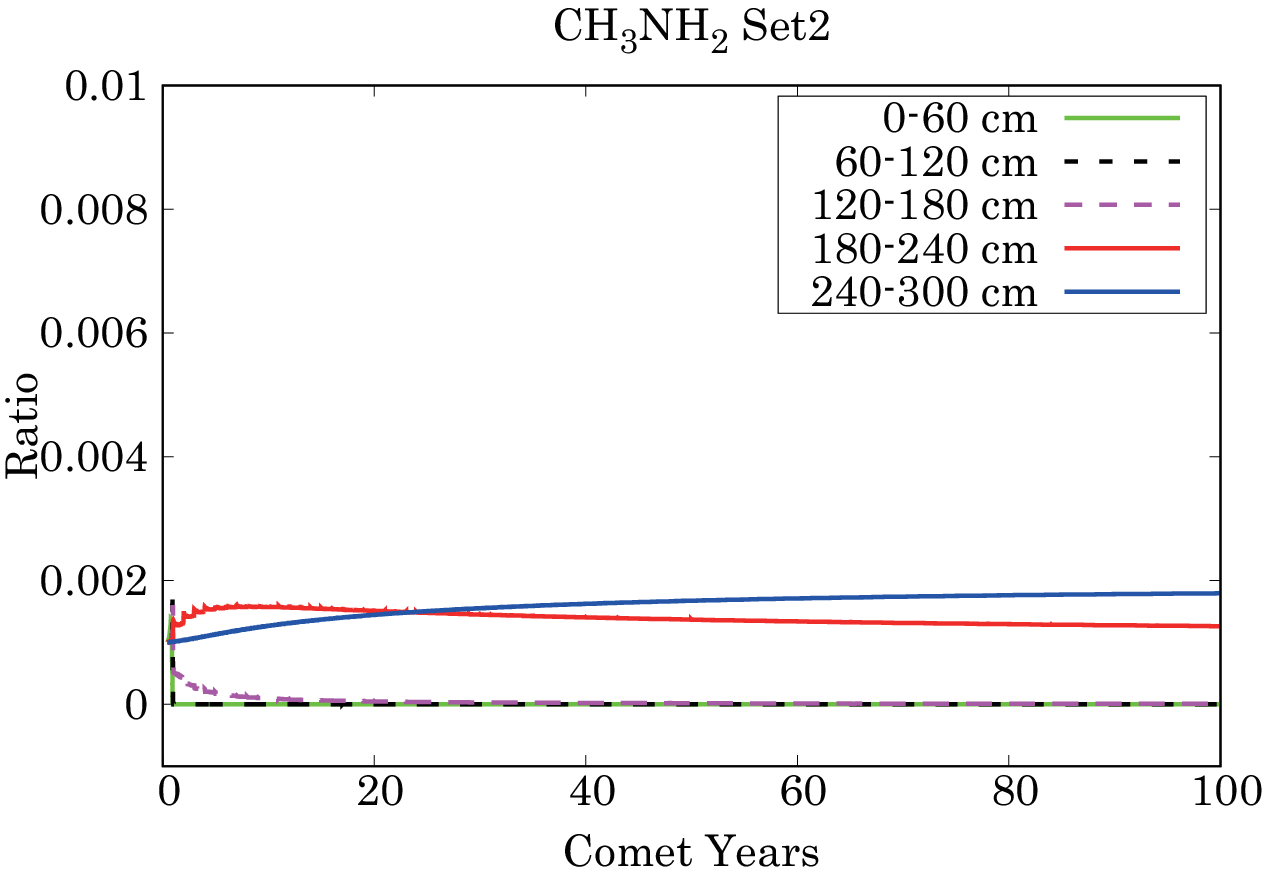}\\
\includegraphics[scale=.65]{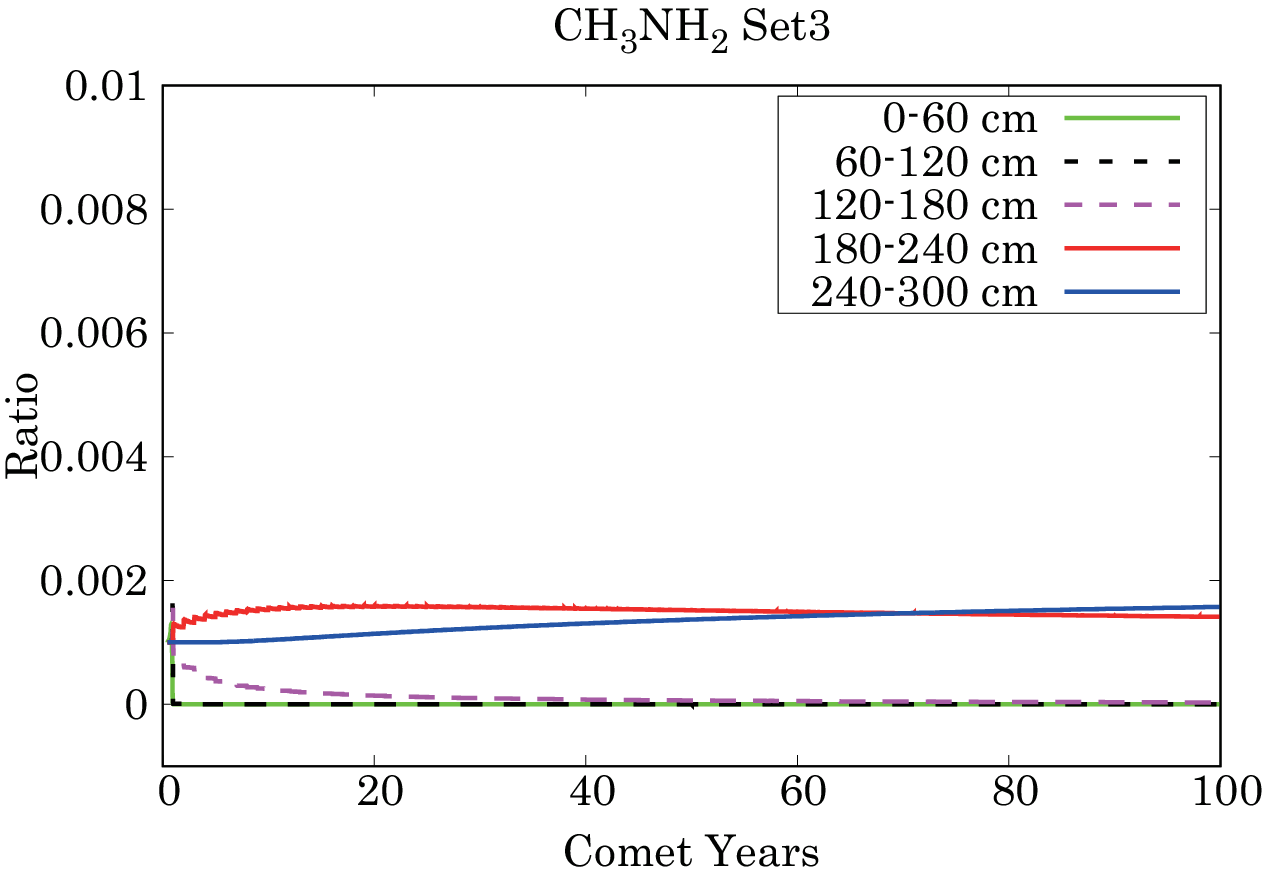}&
\includegraphics[scale=.65]{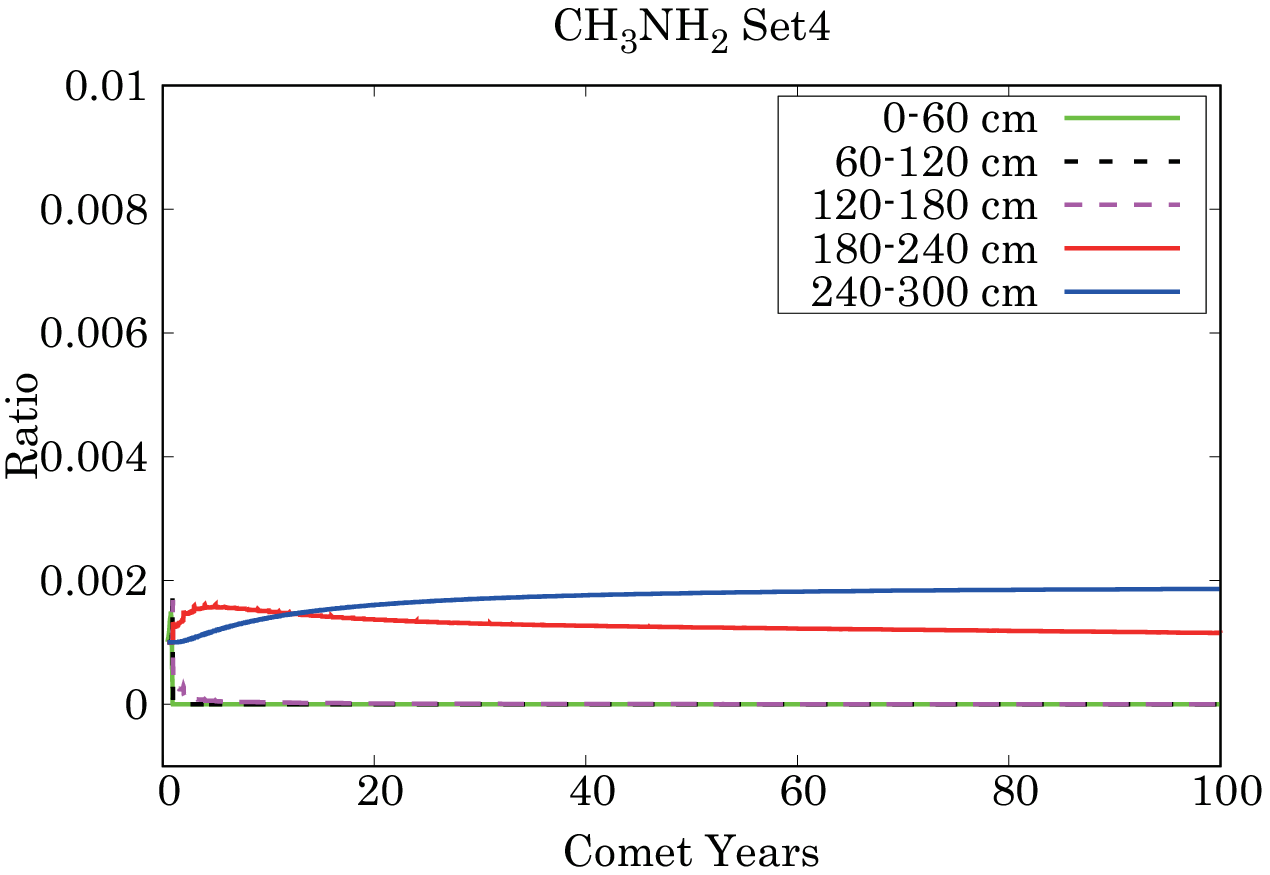}\\
\includegraphics[scale=.65]{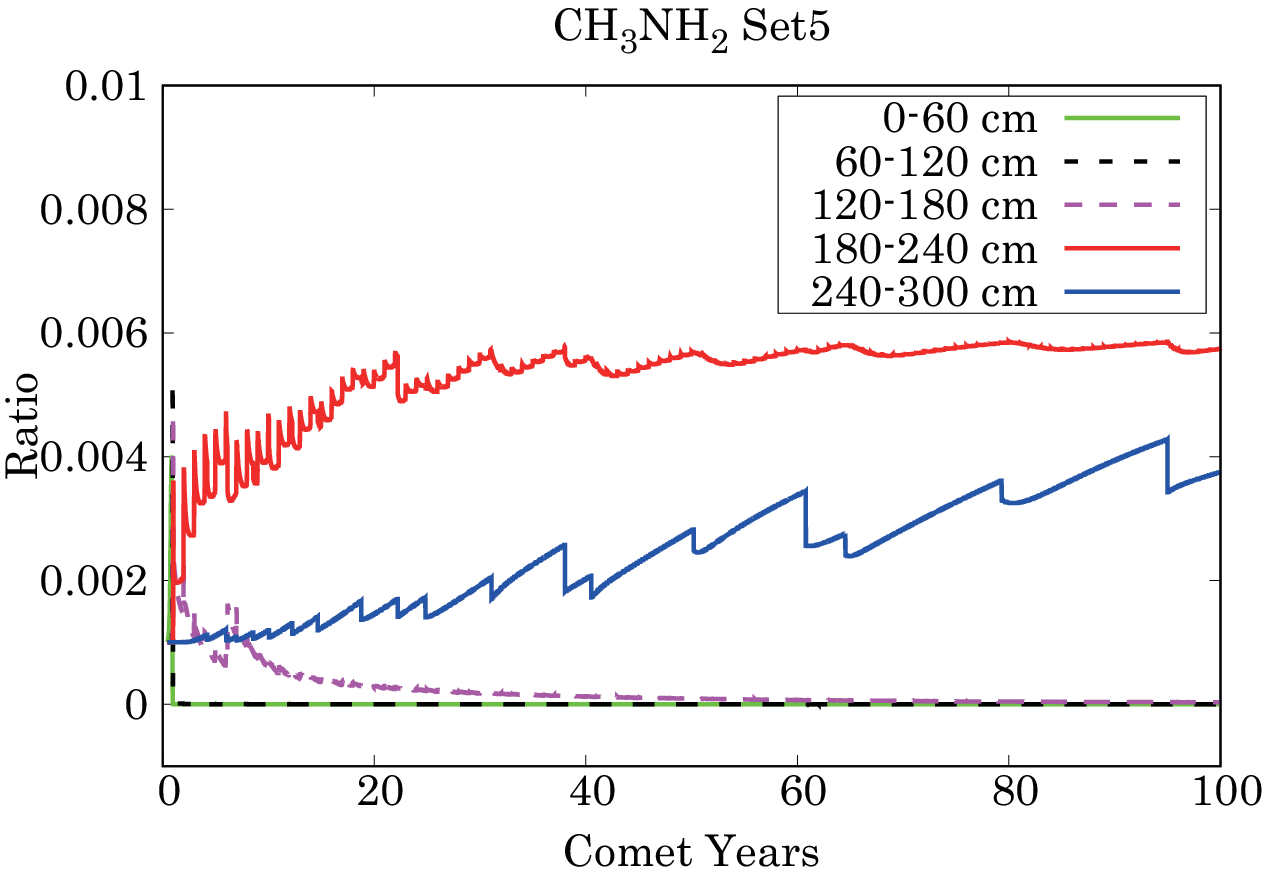}&
\includegraphics[scale=.65]{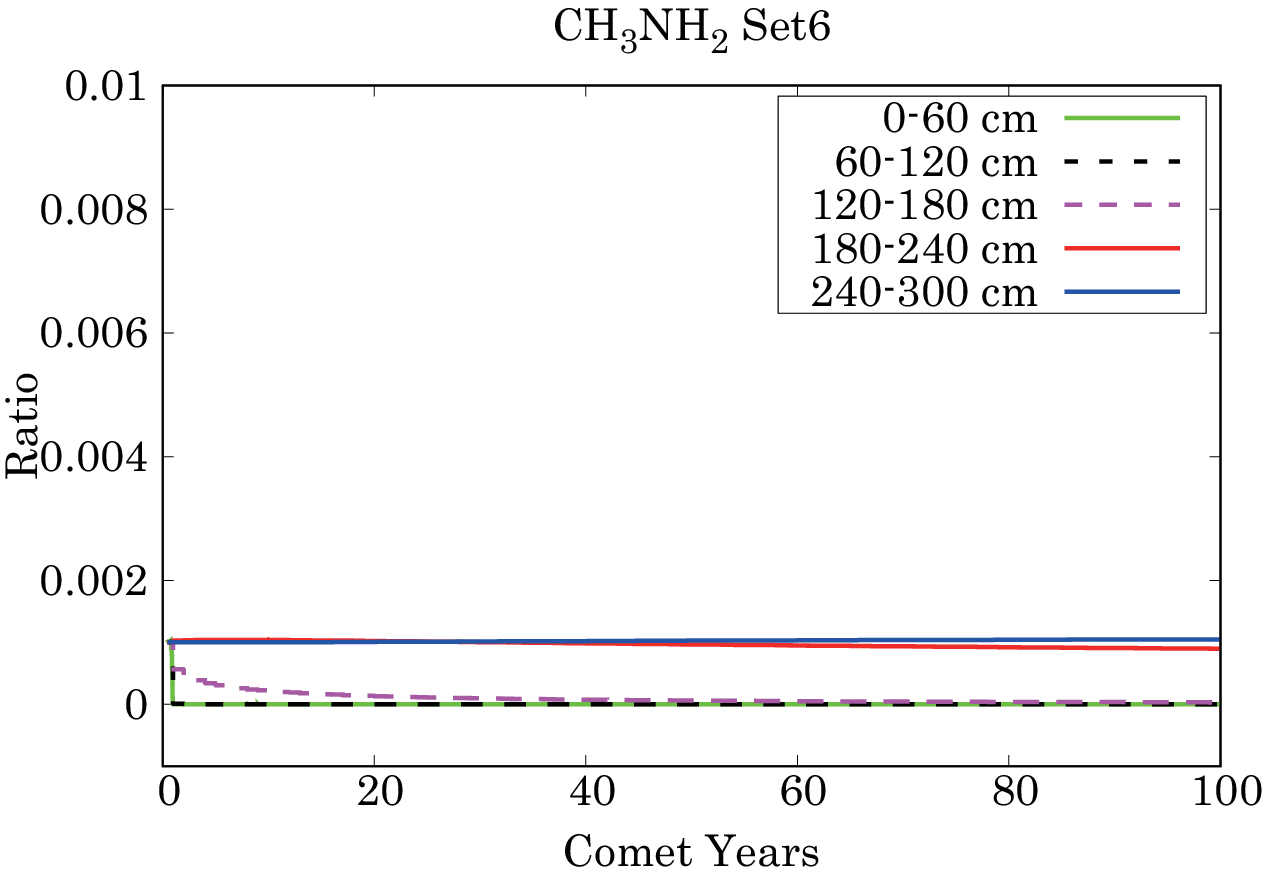}\\
  \end{tabular}
\caption{
\label{fig:CH3NH2}
The same as Figure~\ref{fig:H2O} but with CH$_3$NH$_2$.
}
\end{figure*}

\begin{figure*}
 \begin{tabular}{ll}
\includegraphics[scale=.65]{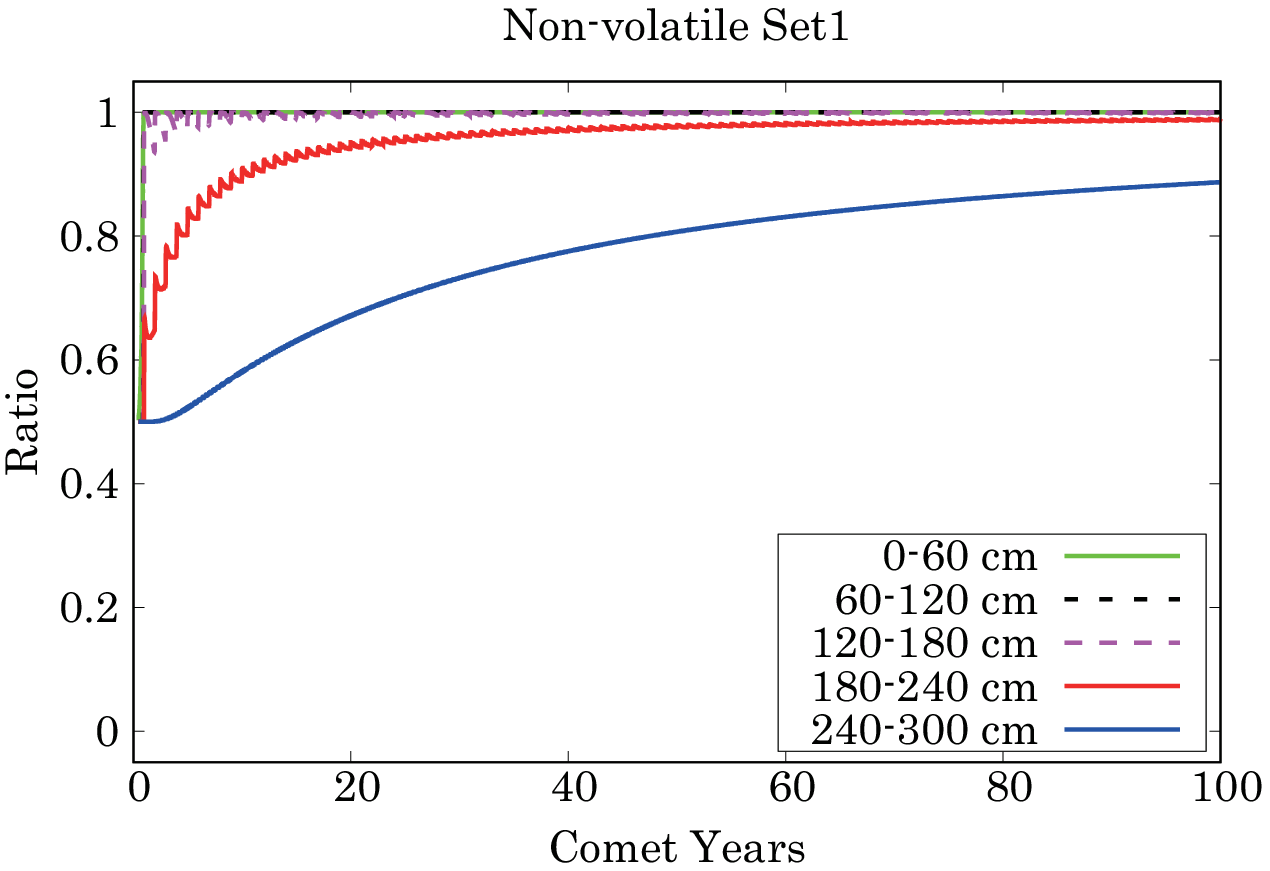}&
\includegraphics[scale=.65]{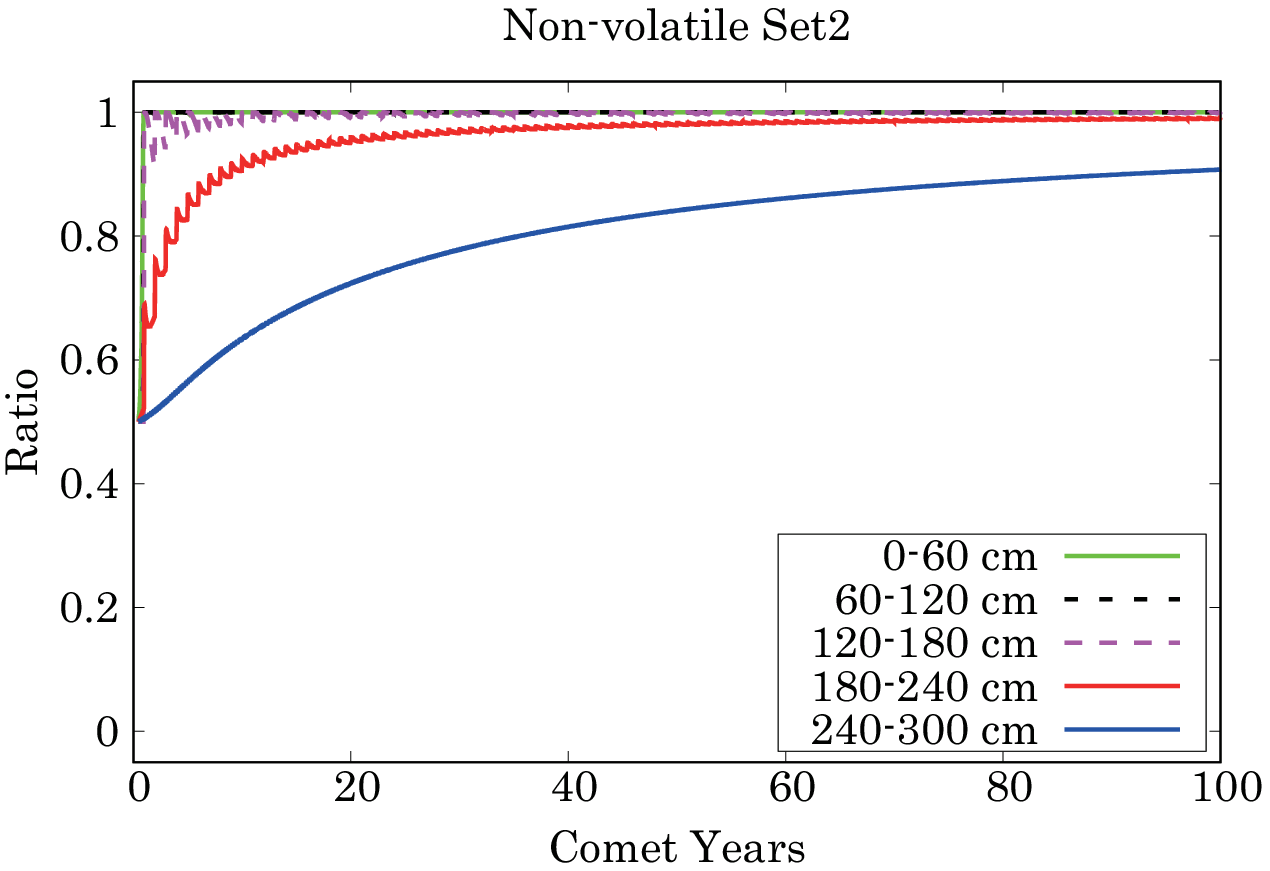}\\
\includegraphics[scale=.65]{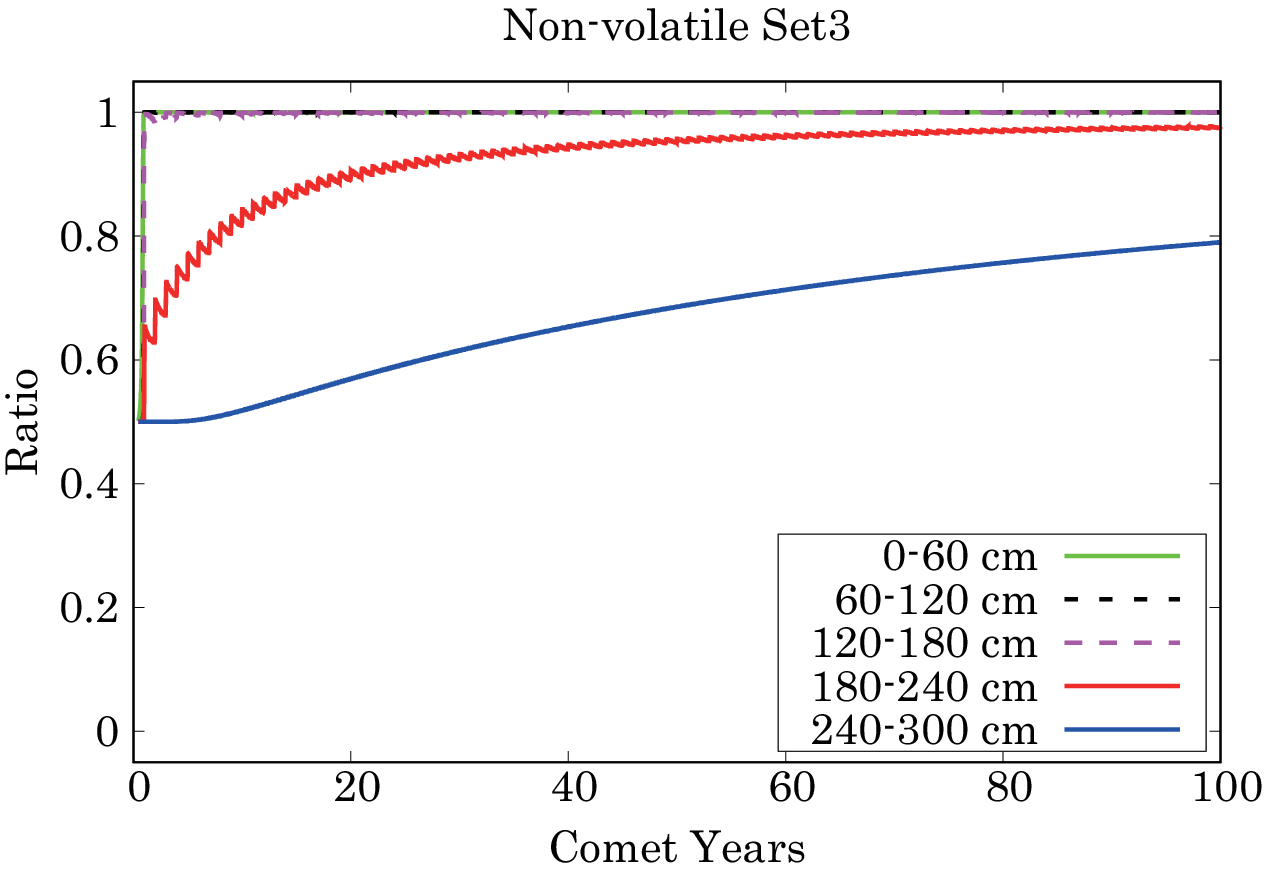}&
\includegraphics[scale=.65]{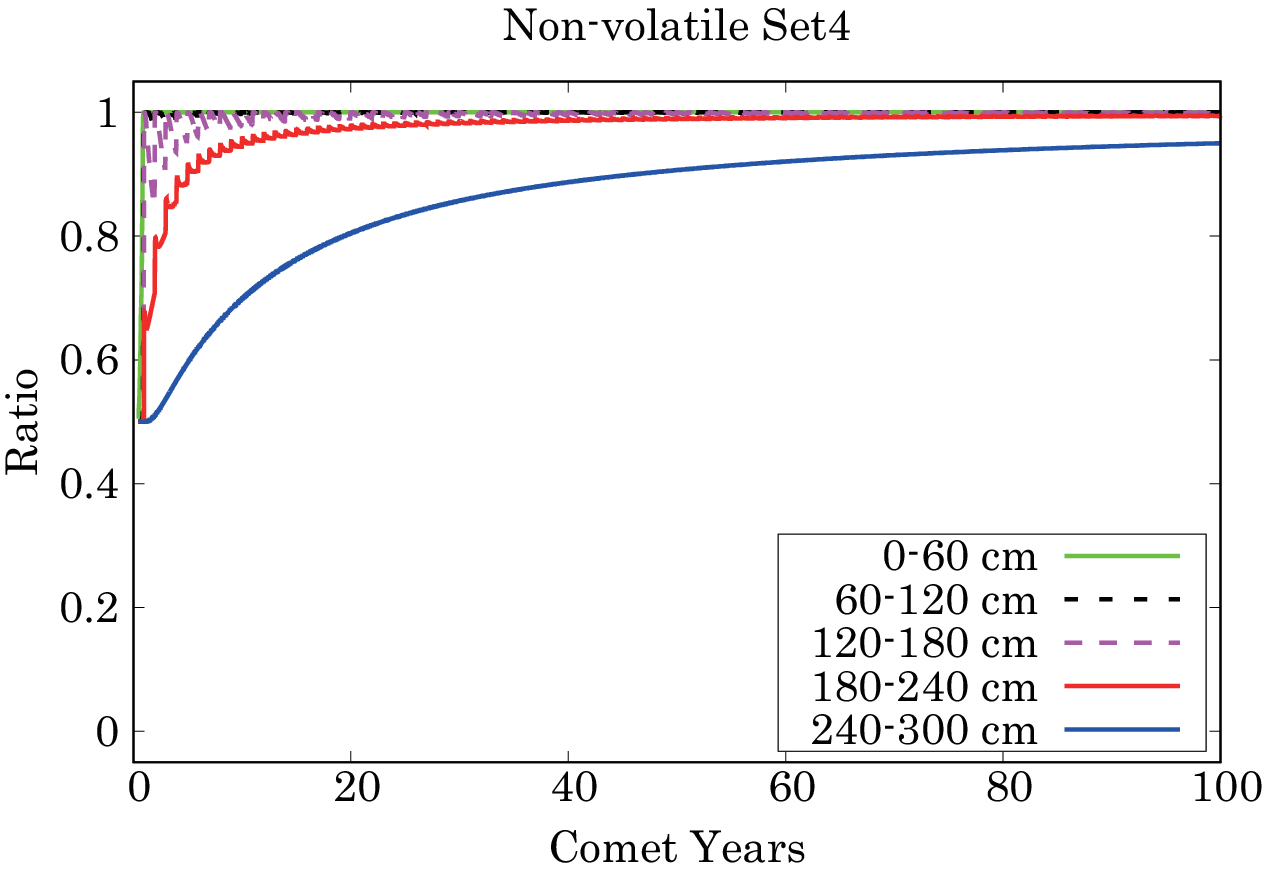}\\
\includegraphics[scale=.65]{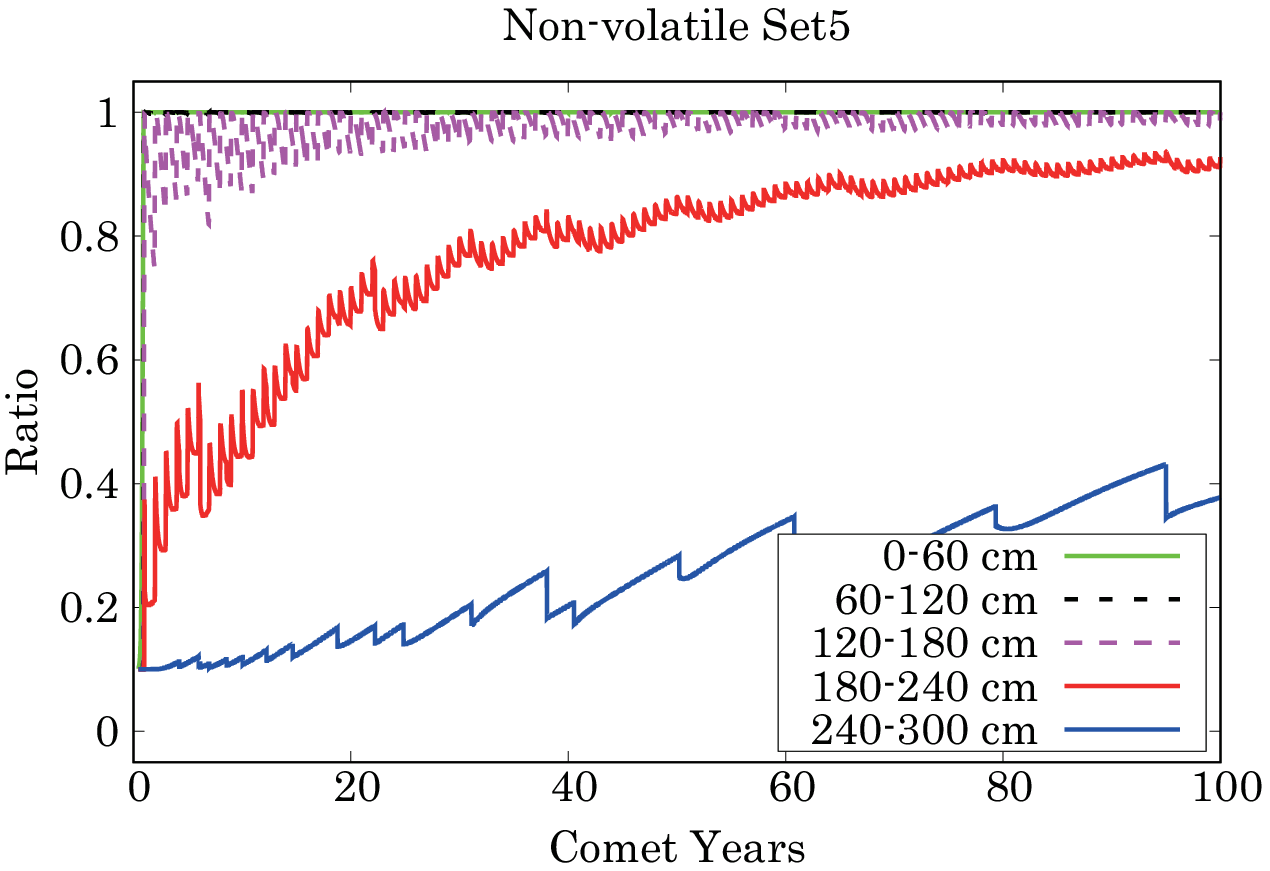}&
\includegraphics[scale=.65]{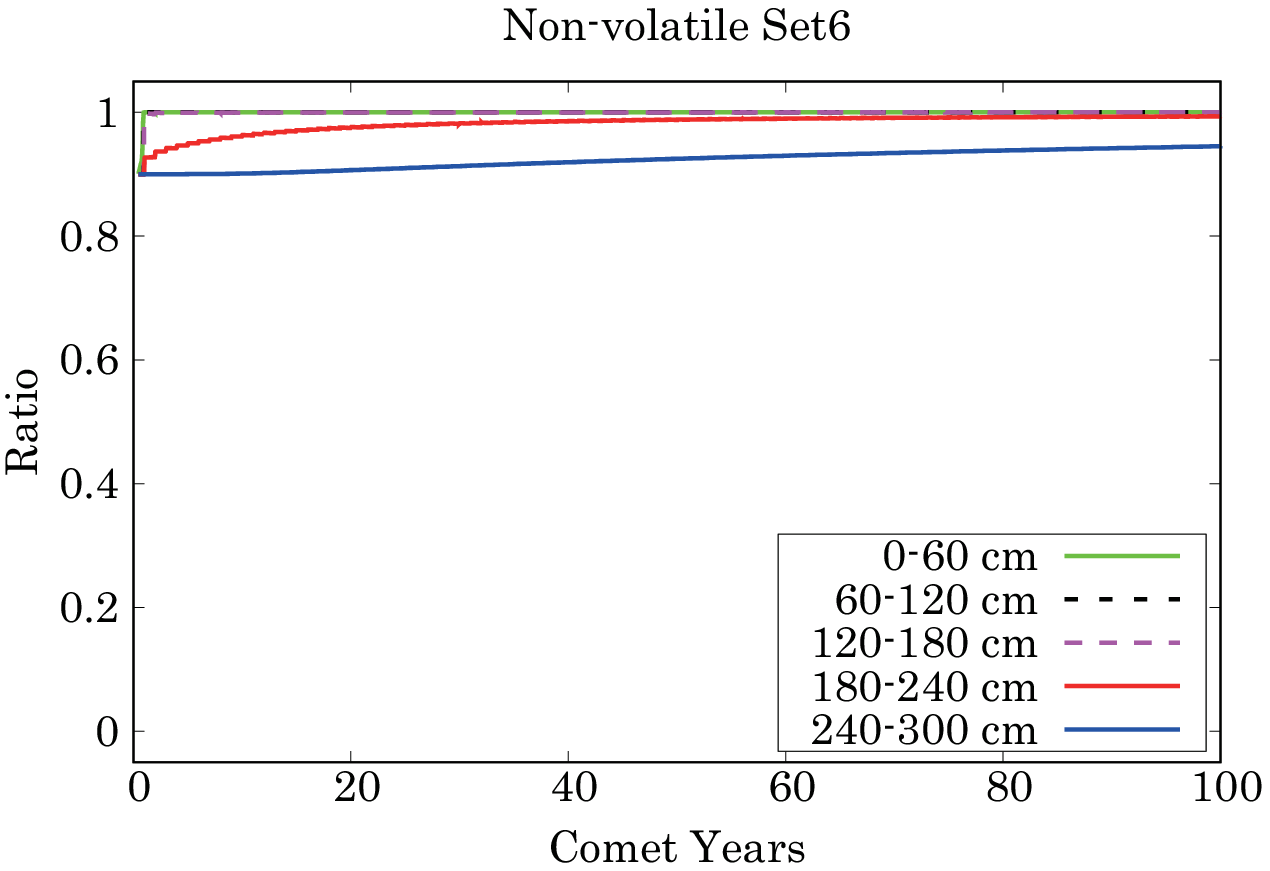}\\
  \end{tabular}
\caption{
\label{fig:non-volatiles}
The same as Figure~\ref{fig:H2O} but with non-volatile species.
}
\end{figure*}

We use Sets~2 to 6 to investigate the parameter dependency of the SCCM.
We change the initial porosity from zero to 0.2 in Set~2 while the other parameters are same as Set~1.
The averaged porosity and the time evolution of the chemical composition of Set~2 (Figures~\ref{fig:porosity} through \ref{fig:non-volatiles}) show a good agreement with Set~1.
Therefore, the initial porosity does not affect our results.

The layer thickness is another parameter of the SCCM model.
We set the layer thickness to be 5 and 20~cm, respectively, in Set~3 and 4, while the other parameters are same as Set~1.
As a result, the time evolution of the porosity and the chemical ratios (Figures~\ref{fig:porosity} through \ref{fig:non-volatiles}) show a small difference, especially in the later phase ($\sim$ 90 comet years) of the simulation.
With these results, we fix the porosity to be zero and the layer thickness to be 10~cm in the following discussion.

\subsection{Simulation Results}
In this section, we discuss the evolution of the comet interior under the assumption of different dust-to-ice ratio.
Since it is difficult to estimate the initial dust-to-ice ratio of the comet before the thermal processing of the Sun, the initial dust-to-ice ratio of SCCM is determined from the observation of the comet 67P.
\cite{Fulle17} suggested a high dust-to-ice ratio of 7.5, while other studies reported a lower dust-to-ice ratio of between 0.6 and 7 \citep[][and references therein]{Choukroun20}.
Therefore, we test three models assuming the different ratios of non-volatile species.
In Sets~1, 5, and 6, respectively, the ratio of non-volatile species is assumed to be 10, 50, and 90$\%$.
We set the layer thickness of 10~cm and the initial porosity of zero for all cases.

In Sets~1, 5, and 6, the averaged porosity in 0-60, 60-120, and 120-180~cm from the surface get to $\sim$0.5, $\sim$0.6, and $\sim$0.1 within 15 comet years (see Figure~\ref{fig:porosity}).
The averaged ratios at the different depth beneath the comet surface are shown in Figures~\ref{fig:H2O} to \ref{fig:non-volatiles}.
At the depths of 0-60 and 60-120~cm, water evaporates so quickly.
The porosity at 0-60 and 60-120~cm is almost 0.5, the ratios of H$_2$O and CH$_3$NH$_2$ are zero, while the ratio of non-volatile species is unity throughout the simulation due to the rapid evaporation of H$_2$O.
The evaporation of H$_2$O leads to the concentration of non-volatile species (Figure~\ref{fig:non-volatiles}) for all sets.
Though the temperature is slightly high for the upper layer, the porosity is small for the deeper layer.
As a result, H$_2$O evaporates from the deeper layer and is frozen on the upper layer when the comet is away from the Sun, which increases the H$_2$O ratio while it decreases the porosity and the non-volatile species ratio.
On the other hand, H$_2$O ratio decreases when the comet approaches the Sun.
This behavior causes the oscillation in the porosity and the ratios.
In addition, the dynamic time evolution of the porosity and ratios are seen in Set~5, where the initial ratio of H$_2$O is 0.9.
The complete evaporation of H$_2$O will lead to porosity of more than 0.7, the merging process of layers leads to a sudden change of the porosity and the ratios.
On the other hand, the merging process is not essential for Set~6, where the ratio of H$_2$O is only 0.1.
With this small ratio, the evaporation of H$_2$O has less significance for the merging process.
The porosity increases slowly once the ratio of non-volatile species gets high.

The layers beneath more than 180~cm take more time to lose H$_2$O due to the lower temperature.
The averaged porosity at 240-300~cm position (the twenty-fourth to thirty-th layer) still changes even after 100 comet years.
The evaporation of H$_2$O is still triggering the merging process in this region.
The difference of CH$_3$NH$_2$ ratio in Sets~1 and 5 at the 240-300~cm position is due to the higher abundance of non-volatile species in Set~1.

\section{Discussion}
\subsection{Water Production Rate}
Since water is the dominant compound of comets, the production rate of water would be the best benchmark of SCCM.
In our model, the production rate of j$^{th}$ species at the time of $t$, Production$[j, t]$, is given by summing the all volume amount of evaporation from the surface ($n=1$) and the indirect evaporation from the deeper layer.
\begin{equation}
\begin{split}
{\rm Production}[j, t] = \rho \times S \times (\frac{\Delta V_{\rm evap}[n=1, j, t]}{\Delta t} \\+ \Sigma_{n\geqq1} \frac{\Delta V_{\rm direct}[n\geqq1, j, t]}{\Delta t}) \times d~{\rm kg}~{\rm s}^{-1},
\end{split}
\end{equation}

where $\rho$ is the mean density of comet 67P, corresponding to 470~kg~m$^{-3}$ \citep{Sierks15}, $d$ is the layer thickness, and $S$ is the total surface area of the comet surface.
Assuming that 67P is the sphere with a radius of 2~km, $S$ is 5 $\times$ 10$^7$ ~m$^2$.
We calculate the water production rate of Sets~1, 5, and 6 at 100 comet years to compare SCCM results with the result of the Rosetta mission.
We show the water production rate of Sets~1, 5, and 6 in Figure~\ref{fig:production_rate} by the function of the distance to the Sun.
The water production rate in August 2014 was 1.2 kg s$^{-1}$ at 3.6~au \citep{Gulkis15}, while the maximum value is 1.73$\times$10$^{2}$ kg s$^{-1}$ after perihelion \citep{Bertaux15}.

The peak production rates of water in Sets~5, 1, and 6 are 447, 73, and 7.7~kg s$^{-1}$, while they are 11, 1.7, and 0.2 kg s$^{-1}$ at 3.6~AU.
The result of Set~6 underestimates the production rate.
The production rate of Set~1 shows a good agreement with the observation.

It would be worth comparing our model with the previous theoretical modeling study of \cite{Keller15}.
They simulated the detailed morphology of 67P, but the thickness of the layer of non-volatile species was assumed to be the micrometer scale.
According to their model~B, the water production rate at 3.6~AU was 61~kg s$^{-1}$, which is higher than the observation by \cite{Gulkis15} by order of magnitude.
Our agreement on the water production rate emphasizes the validity of SCCM.
We evaluated the erosion depth with our model by summing up the production rate along with an orbit followed by the division by the density.
\begin{equation}
{\rm Erosion}[j] = \Sigma_{t} \frac{{\rm Production} [j, t] \Delta t}{\rho}~{\rm m}
\end{equation}
We find the erosion depths of 0.3, 0.03, and 0.006~m for Sets~1, 5, and 6, respectively, at 100 comet years.
These depths are smaller than the result of \cite{Keller15}, where the erosion depths were between 0.67 and 2.9.

\subsection{Chemical Composition}
We compare the water production rates obtained by Equation (14) with the observation of the Rosetta mission.
The production rate ratio, CH$_3$NH$_2$/H$_2$O, is 0.007, 0.004, and 0.002 for Sets~1, 5, and 6, respectively, when the distance to the Sun is more than 2~AU.
These ratios agree well with the observed ratio of 0.006 \citep{Goesmann15}.

\subsection{Timescale of the Heating}
\begin{figure*}
 \begin{tabular}{ll}
\includegraphics[scale=.65]{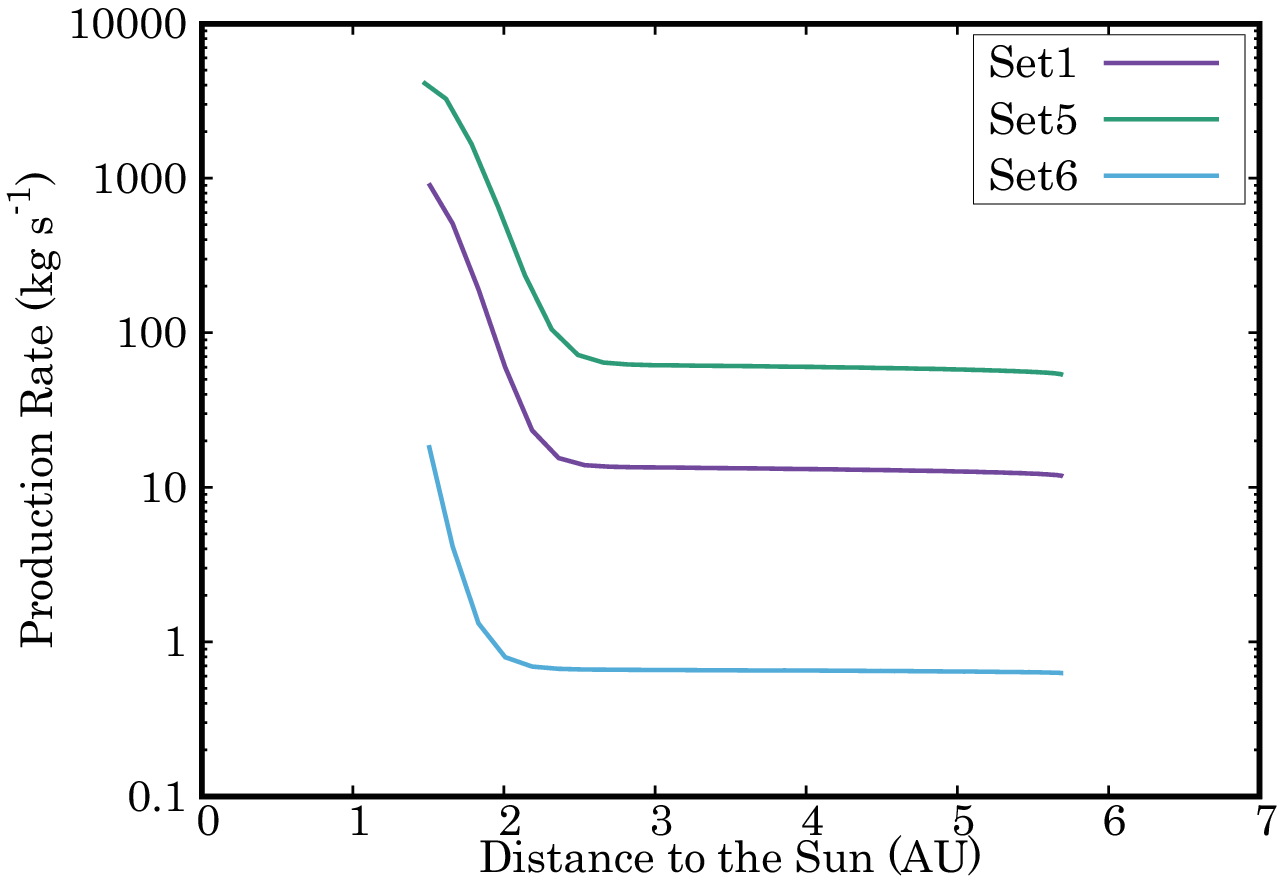}&
\includegraphics[scale=.65]{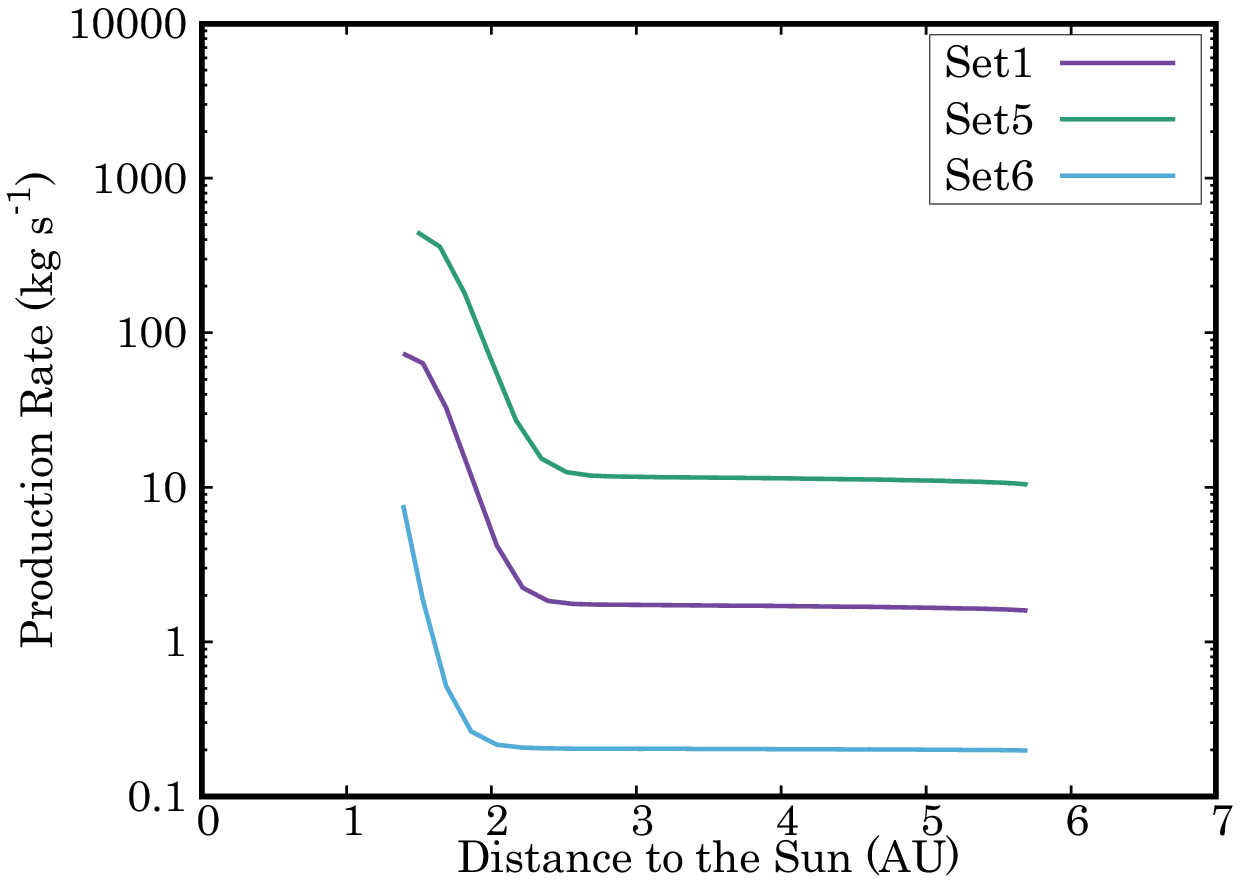}\\
  \end{tabular}
\caption{
\label{fig:production_rate}
(Right) The water production for Sets~1, 5, and 6, 100 comet years are depicted with the different scale of y-axis.
This case simulate the thermal processing of the surface of 67P for the long timescale beyond 1959.
(Left) The water production for Sets~1, 5, and 6, 7-8 comet years are depicted with the different scale of y-axis.
This case represent the specific case where 67P did not experience the hearing before 1959.
}
\end{figure*}

If comet 67P did not experience any heating before 1959, the heating timescale of 67P would be only eight comet years. 
We show the water production rates at this age with Sets~1, 5, and 6 in Figure~\ref{fig:production_rate}.
At this early phase of the thermal processing, the water production rates for Sets 1 and 5 are as high as $\sim$10$^{3}$ kg s$^{-1}$, since there is abundant water near the comet surface.
The water production rate for Set~6 is $\sim$10 for the perihelion, while it is $\sim$0.6 at 3~au.
Therefore, the SCCM model can explain the water production rate if we assume that the ratio of the non-volatile species is high for the interior of 67P.
We note that CH$_3$NH$_2$/H$_2$O ratio does not dramatically change with the history of the thermal processing.
Without any information of H$_2$O ratio inside 67P, it would be difficult to constrain its history of thermal processing.

\subsection{Possible Effect to the Chemistry in the Coma}
The comet activity may release the concentrated materials to the coma.
Rosetta observed 34 outbursts within three months surrounding its perihelion passage, suggesting this event is ubiquitous at least for 67P \citep{Vincent16}.
Since the outburst is defined by the observable change of the luminosity, smaller activity should be more ubiquitous.
Several mechanisms for comet activity would be crystallization of amorphous ice, enrichment of volatile inside the comet, and receding fractured cliffs \citep[e.g.,][]{Agarwal17,Vincent16}.

\cite{Altwegg16} suggested the correlation of glycine abundance with the dust density, giving the possibility that glycine is not directly originated from the nucleus of the comet but evaporated from dust particles, ejected by comet activity.
The dust particles are efficiently heated by solar flux \citep{Lien90}, resulting in a higher temperature to release the absorbed glycine.
\cite{Hadraoui19} also showed that glycine could originate from icy dust particles with water ejected from the nucleus.
Though these studies does not consider the concentration of the non-volatile species, the release of non-volatile species from the comet surface can affect the observed molecular abundances in the coma.

\subsection{Effect to the Early Earth Chemistry}
It is believed that the comets would have hit the early Earth to provide the materials for the origin of life \citep{Ehrenfreund02}.
Concentration of materials, perhaps in micrometeorites of cometary origin, would provide the unique chemical condition to the volcanic hot spring pools to discuss the chemical evolution toward life.
However, the surviving rate of organic materials during the cometary bombardment, especially for the large comets, is still controversial.
\cite{Pierazzo99} investigated the thermal decomposition rates of amino acids based on the time evolution of the temperature after the impact with their hydrocode simulations.
They showed that the time evolution of the temperature after the impact strongly depends on the position.
The opposite side of the impact position is less heated and provides a higher possibility of survival of amino acids.

Figure~\ref{fig:porosity} shows that the timescale of the concentration process for 67P is very short.
The concentration of non-volatile species at depths between 0 and 100~cm would be achieved within ten comet years, which corresponds to 65.7~years.
As we can see the concentration process for the deeper layers in Figure~\ref{fig:porosity}, the typical depth of this concentration would be several meters beneath the surface.
Since this timescale is very short for the planet formation, most comets are expected to hit the early Earth after the concentration of non-volatile species.
The non-volatile species would include amino acids and other complex and large molecules, which may be essential for the chemical evolution toward the origin of life.
Also, the concentration of these materials at the outer layer would show a high survival rate after the cometary impact due to the lower temperature compared to the impact position.
The outer layers on the opposite side of the comet would be interesting for further studies.

\section{Summary}
We investigated the evolution of the composition of the outer cometary layers after the heat processing by the Sun. 
We developed SCCM model to see the importance of the concentration process of non-volatile species by simulating the evaporation and the merging processes along with the orbit of comet 67P.
We found that the SCCM model successfully reproduced the water production rate and CH$_3$NH$_2$/H$_2$O ratio for 67P.
SCCM model suggests that non-volatile species may concentrate quickly near the comet surface, while the initial ratio is saved at the deep inside of comets.
Comets may have a layer of non-volatile species beneath the several meters from the surface when they hit the early Earth.
The outer layer of the comet would be interesting for further studies as they give the unique chemical condition to the volcanic hot spring pools.

\begin{acknowledgements}
We would like to thank the anonymous referee for constructive comments that helped to improve the manuscript. We are grateful to Dr. Takefumi Ootsubo and Dr. Hideyo Kawakita for the discussion.
This work was supported by JSPS KAKENHI Grant Number JP16J09590 and 15J10864.
\end{acknowledgements}

%
%
\end{document}